\documentstyle[11pt,aaspp4]{article}

\slugcomment{Submitted to Ap. J.}

\lefthead{Ivanov et al.}
\righthead{Testing the AGN-Starburst Connection in Seyfert Galaxies
}

\begin{document}

\title{Testing the AGN-Starburst Connection in Seyfert Galaxies.}

\author{Valentin D. Ivanov, George H. Rieke, Christopher E. Groppi,
Almudena Alonso-Herrero\footnote{Current address: Department of 
Physical Sciences, University of Hertfordshire, College Lane, 
Hatfield AL10 9AB, England; aalonso@star.herts.ac.uk}, Marcia J. 
Rieke, C. W. Engelbracht,}
\affil{Steward Observatory, The University of Arizona, Tucson, AZ 85721,\\
        vdivanov, grieke, cgroppi, aalonso, mrieke, \&
        cengelbracht@as.arizona.edu}

\begin{abstract}
We use the CO band at $2.3\,\mu$m to constrain the populations 
of young stars in the central regions of Seyfert galaxies.
We report new CO band spectroscopy of 46 Seyfert galaxies. In most cases,
the observed CO indices appear diluted by the presence of a non-stellar
component (most likely, warm dust surrounding the active nucleus). We
used $\rm JHKL$ aperture photometry to estimate the non-stellar
contribution at $2.3\micron$. We successfully corrected the CO band for
the dilution for 16 galaxies which were not dominated by the non-stellar
component. Comparing with CO indices measured in elliptical and purely
starbursting galaxies, we find no evidence for strong starbursts in the
majority of these galaxies.

\end{abstract}


\section{Introduction}

There are many observational examples of starbursts occurring in active
galaxies (e.g. Mauder et al. 1994\markcite{mau94}; Maiolino, Rieke \&
Keller 1995\markcite{mai95}; Heckman et al., 1997\markcite{hec97};
Gonzalez-Delgado et al., 1998\markcite{gon98}; Nelson \& Whittle,
1998\markcite{nel98}; Kunth \& Contini, 1999\markcite{kun99}). As a
result, possible connections between active galaxies and circumnuclear
starbursts are the subject of extensive speculation.

Such associations might arise just because
circumnuclear starbursts have a common cause with
AGNs. For example, mergers, interactions between galaxies,
and bars (Shlosman, Frank \& Begelman, 1989\markcite{shl89};
Shlosman, Begelman \& Frank, 1990\markcite{shl90};
Maiolino et al., 1997\markcite{mai97};
Knapen, Shlosman \& Peletier, 1999\markcite{kna99}) have
all been invoked as a means to distort the
gravitational potential of the host galaxy, allowing
the gas to lose angular momentum to the stars,
so it can fall inward toward the AGN. All of these phenomena appear
also to be associated with nuclear starbursting.

A more direct connection has been proposed by von Linden et al.
(1993\markcite{von93}),
who suggest that starburst-induced turbulence in the ISM
is responsible for the final stages of infall of gas into
the nuclear engine. Weedman (1983)\markcite{wee83} pioneered
an AGN model involving a large number of small accretors -
the remnants of the massive stars in a compact
circumnuclear starburst, preceding the AGN stage.
Bailey (1980)\markcite{bai80},
David (1987a, 1987b)\markcite{dav87a}\markcite{dav87b}, and
Norman \& Scoville (1988)\markcite{nor88} refined
this idea, pointing out that a compact star cluster can
form a super massive black hole. The theoretical
understanding of this process was advanced by the
efforts of Perry \& Dyson (1985)\markcite{per85},
Perry (1992)\markcite{per92}, and more recently, of
Williams, Baker \& Perry (1999)\markcite{wil99}.

It has also been proposed that a starburst can mimic an AGN
(Terlevich \& Melnick, 1985)\markcite{ter85}.
Terlevich et al. (1992)\markcite{ter92} showed that at least low
luminosity Seyfert 2's can be explained via active circumnuclear
star formation alone. However, this model has difficulties
explaining the rapid X-ray variability, and the broad iron
emission lines of more luminous active galaxies.
Cid Fernandes (1997)\markcite{cid97}
summarized the current status of the model and suggested
a dichotomy among AGNs, in which some are powered
by a black hole and others by a nuclear starburst.

Still another possibility is that circumnuclear
starbursts strongly alter the observed properties
of AGNs. For example, Ohsuga \& Umemura (1999)\markcite{ohs99} consider
a model in which the wind from the circumnuclear
starburst forms an obscuring dust wall. The alternative
process, in which an AGN disturbs the galactic ISM
and triggers a starburst was reported by
van Breugel et al. (1985)\markcite{van85},
van Breugel \& Dey (1993)\markcite{van93}, and
Dey et al. (1997)\markcite{dey97}.
This behavior is usually related to the expansion of a radio source into the
ambient medium. Also, once formed, by whatever means,
the black hole can gravitationally disturb the interstellar gas,
triggering a starburst.

A better understanding of the frequency of AGN
and starburst coincidences is required to probe
the connection between the two phenomena.
A conventional method to search for young starbursts is to look for
stellar absorption features in the ultraviolet spectra of galaxies
(e.g. Heckman et al., 1997\markcite{hec97};
Gonzalez-Delgado et al., 1998\markcite{gon98}).
Current instrumentation allows this approach to be
applied in just a few cases; Heckman (1999)\markcite{hec99} found
only four Seyfert 2 nuclei bright enough to
obtain UV spectra with sufficiently high
signal-to-noise to identify young stars. It is
possible that they show starburst
signatures because of the selection bias toward galaxies with high
surface brightness, which could be
a result of the active star formation.
In addition, application of this technique to AGNs
is restricted by the extinction. Since the ISM lies around the
AGN in a complex three-dimensional distribution, a significant component
of the observed UV flux may come from the stellar population in front
of the active nucleus and not be associated closely with it.

In the optical, Kunth \& Contini (1999)\markcite{kun99} detected
emission signatures of Wolf-Rayet stars in the spectra of two
Seyfert 2 galaxies out of five observed.
CaII triplet absorption at $\rm 8555\AA$ is strong in
the spectra of massive stars and can also
indicate the presence of starbursts
(Terlevich, Diaz \& Terlevich, 1989\markcite{ter89};
Terlevich, Diaz \& Terlevich, 1990\markcite{ter90};
Forbes, Boisson \& Ward, 1992\markcite{for92};
Nelson \& Whittle, 1998\markcite{nel98}). However, 
studies of this line in Seyfert galaxies were not
conclusive because for a fraction of these galaxies the
CaII index values were inconsistent with a normal galaxy spectrum
diluted by a featureless power law continuum (see also
Persson, 1988\markcite{per88};
Diaz, Terlevich \& Terlevich, 1989\markcite{dia89}).

The infrared range has an advantage in probing AGN environments 
because of the lower extinction.
Since the strengths of the CO bands at $1.6\micron$ and $2.3\micron$
are sensitive to the stellar surface gravity, they can
indicate the presence of young red supergiants. The CO band
strength is also a function of metallicity. However, in the bulges of
large spirals, the typical haunt of a Seyfert nucleus, the surface
gravity dependence dominates. Empirically, quiescent spirals have a
very narrow range of $2.3\micron$ band strengths (Frogel et al., 1978;
Frogel, 1985\markcite{fro85}) and most starburst galaxies have bands
substantially deeper than this value (Ridgway, Wynn-Williams \& Becklin
1994\markcite{rid94}; Goldader et al. 1995; Engelbracht 1997)

Although the $2.3\micron$ strength has proven useful
in identifying and studying starbursts, in
applying similar techniques to Seyfert galaxies one
must contend with the possibility that the band
can be diluted by radiation from the warm dust surrounding
the central engine in the AGN. Oliva et al. (1995)\markcite{oli95}
therefore used the second overtone
at $1.6\micron$ to probe the environments of
AGNs. Unfortunately, this band is weak and its dependence on
metallicity is strong comparable to $2.3\mu$m CO band.
To avoid these issues, Oliva et al. (1999)\markcite{oli99}
derived the mass to luminosity of the nuclear region
and identified starburst/Seyfert combinations where
this ratio was low, rather than relying
on the CO band strengths. Both of these approaches require
high signal to noise spectra of moderately high resolution.
Thus, they can only be applied to AGNs that are bright in the near
infrared, which may result in a selection bias toward
examples with strong circumnuclear starbursts.

To address the Seyfert-starburst connection more thoroughly,
we obtained $\rm K$-band spectra of 46 Seyfert galaxies
in the CfA sample at the $2.3\,\mu$m CO bandhead.
In addition, we obtained spectra of 5 ellipticals,
and used 19 starburst galaxy spectra from 
Engelbracht (1997)\markcite{eng97} for comparison.
With this sample, we hope to alleviate the typical small number
statistics of previous studies, and to overcome the bias
toward galaxies with high surface brightness. We apply
a technique based on $\rm JHKL$ aperture photometry
and imaging to correct for the dilution of the CO band from dust
heated by the AGN. We then compare the corrected
CO band strengths with those of the ellipticals and starbursts.

\section{Observations and Data reduction}

\subsection{Infrared Spectroscopy}

The infrared spectra were taken over a period of four years at three
different telescopes with FSpec (Williams et al., 1993)\markcite{wil93},
a cryogenic long slit near infrared spectrometer utilizing a NICMOS3
256x256 array. We used $R~\approx~700$, $R~\approx~1200$, and
$R~\approx~3000$ resolution modes, corresponding to gratings of 75, 300,
and 600 lines/mm. The slit width was 1.2 arcsec at the MMT, 2.4 and 3.6
arcsec at Steward Observatory 2.3m and 1.55m telescopes. The plate scales
were 0.4, 1.2 and 1.8 arcsec/pixel respectively. The observing log is
presented in Table~\ref{tbl-1}. The galaxies are sorted in order of right
ascension.

\placetable{tbl-1}

The spectra were reduced using IRAF tasks written specifically for
FSpec (Engelbracht, 1999\markcite{eng99}).
An average of sky background images taken
immediately before and after each object image was subtracted to remove
the sky emission lines, and simultaneously, the dark current and offset
level. Dark-subtracted dome flats were
used to correct for pixel-to-pixel sensitivity variations.
Then the object images were shifted and combined to produce a single
two-dimensional spectrum. Known bad pixels were masked out. The large
number of object images allowed us to exclude any remaining bad pixels and
cosmic rays. The one-dimensional spectrum was extracted by fitting a 3-5
order polynomial to the continuum in the two-dimensional image.
The angular sizes of the extraction apertures are given in Table~\ref{tbl-1}.

The object spectra were divided by spectra of stars close to solar type,
observed at the same airmass, and then multiplied by a solar spectrum to
remove the effects of the atmospheric absorption (Maiolino, Rieke \& Rieke,
1996\markcite{mai96}). The true shape of the continuum was restored by
multiplying the spectra by the ratio of black bodies with standard star
and Solar effective temperatures. We used the OH airglow lines (Oliva \&
Origlia, 1992\markcite{oli92}), complemented when necessary with Ne-Kr
comparison lamp spectra, for wavelength calibration.

Special attention was paid to determining the signal-to-noise of our
data. The main sources of uncertainty are the photon statistics for
faint objects, the strong sky emission lines and sky variation. First,
we calculated the photon statistics, taking into account the object, the
sky emission and the contribution from the standard star used for
atmospheric correction. This is an optimistic estimate since it neglects
the sky variations. Second, we calculated the noise from a relatively
featureless part of the spectrum, which gives a pessimistic estimate
because weak absorption lines were included as noise. Finally, for a
few galaxies we calculated the signal-to-noise in the process of
combining one-dimensional spectra extracted from each individual
two-dimensional image. This is the most robust method and it accounts
for all possible sources of error. The last two determinations gave
virtually the same answers. The second technique is easier to implement
but it has the disadvantage of estimating the noise level only on a
clear continuum part of the spectra. To obtain the noise level at the
lines and bands of interest, we rescaled the photon statistics noise
spectrum to match the noise determined from the clear part of the
spectra. The corresponding averaged signal-to-noise ratio is shown in
Table~\ref{tbl-1}.

\subsection{Infrared Line Measurements}

The line measurements used Gaussian fitting. 
Broadband $\rm K$ images, kindly provided by
McLeod \& Rieke (1995\markcite{mcl95}), were used to flux calibrate the
spectra using the absolute calibration of Campins, Rieke \& Lebofsky
(1985)\markcite{cam85}. In turn, most of the images
were re-calibrated using our
aperture photometry (discussed below). The spectra in Figure~\ref{fig1}
are divided by a low-order polynomial fit to the continuum, and shifted 
vertically, for display purposes. The individual absorption feature 
measurements are listed in Tables~\ref{tbl-2}, and~\ref{tbl-3}. 
Emission lines will be discussed in a separate paper.
$1\sigma$ errors for the flux measurements include uncertainties both
in the spectra (sky variation, continuum placement, shot noise) and in
the flux calibration. $1\sigma$ errors for the equivalent line widths
include only the uncertainties in the spectra which are dominated by
the continuum placement error. The latter was estimated by taking
measurements at different continuum positions. Empty items in the
tables denote a lack of spectral coverage, and zero equivalent widths
stand for a non-detection. In the latter case, the $1\sigma$ error has
the meaning of a $1\sigma$ upper limit.

\placetable{tbl-2}
\placetable{tbl-3}
\placefigure{fig1}

\subsection{CO Band Measurements}

Our high resolution spectra lack the wavelength coverage to measure the
photometric CO index of Frogel et al. (1978\markcite{fro78}) or
the spectroscopic CO index of Doyon, Joseph \& Wright
(1994\markcite{doy94}). We defined a narrower CO index to
utilize the complete set of data. Our index has the advantage of
being insensitive to extinction. It follows the
recipe of Doyon, Joseph \& Wright (1994)\markcite{doy94}, but the
measured regions are narrower and closer:
\begin{equation}
\rm CO~=~-2.5~\times~log_{10}(F_{22980\AA}/F_{22850\AA})
\end{equation}
where $\rm F_{22980\AA}$ is flux averaged in the band within a
$\rm 100\AA$ wide window centered at $\rm 22980\AA$, and
$\rm F_{22850\AA}$ is flux averaged on nearby continuum within a
$\rm 100\AA$ wide window centered at $\rm 22850\AA$, in the rest frame
of the object.

A narrow index that does not encompass totally the spectral feature
is sensitive to velocity dispersion variations. To alleviate this
problem, we left out NGC~5990 and NGC~6240 from the comparison starburst
sample, since they are known to have exceptionally large velocity
dispersions. NGC~6240 is also not a typical starburst galaxy, since its
faint $Br\gamma$ is indicative of a relatively old burst of star
formation.

An additional incentive to use narrower bandpasses is related to
possible uncertainties in the continuum shape of the infrared spectra.
The CO feature is close to the edge of an atmospheric window and we
cannot measure the continuum level on both sides of the spectral
feature. Although during the data reduction the original continuum
shape was restored after the spectra were multiplied by the solar
spectrum, we could not exclude some small deviations in the continuum
shape. The narrow CO index minimizes the effect of the continuum shape
on our analysis. To quantify this effect we used our $\rm JK$
aperture photometry to calculate the continuum slope at $2.3\micron$
via simple linear interpolation. We imposed this slope on continuum
divided spectrum and measured the CO index again.

The statistical errors in the CO index 
$(\sigma_{CO}=0.01$ mag for a spectrum with S/N=30 per pixel) are
negligible, compared to the systematic errors due to the uncertain
continuum slope. To determine the true uncertainties, we took advantage
of the multiple spectra of some galaxies, taken on different nights,
and with different spectral resolution. We found the typical total
uncertainty in the CO index to be 0.02 mag. 

Finally, we compared our CO index with that of Doyon, Joseph \& Wright
(1994)\markcite{doy94} calculating both indices for the set of stellar
spectra by Kleinmann \& Hall (1986)\markcite{kle86}. For 26 stars, the
linear conversion is:
\begin{equation}
\rm CO(this~work)~=~(0.97\pm0.03)~\times~CO(Doyon)~+~(0.002\pm0.006)
\end{equation}
with standard deviation of residuals of 0.02. That is, within the errors
the two indices are equivalent. Our CO index is related to other systems
of CO indices via transformations derived by Doyon, Joseph \& Wright
(1994)\markcite{doy94}.

\subsection{Aperture Photometry}

The aperture photometry was carried out at the MMT using the
facility single pixel InSb aperture photometer. Aperture diameters,
fluxes and observed colors with $1\sigma$ errors are shown in
Table~\ref{tbl-4}. The calibration is based on the
standard star network of Elias et al. (1982\markcite{eli82})
with zero points based on Campins, Rieke \& Lebofsky
(1985\markcite{cam85}). With data already available in the
literature, a full set (JHKLN) of ground-based photometry is
now available for the entire CfA sample.

\placetable{tbl-4}

\subsection{$\rm K$-band Imaging}

$\rm K$-band imaging was taken from the survey of McLeod \&
Rieke (1995)\markcite{mcl95}. All the images were calibrated
using our aperture photometry. Many of the images had
independent calibrations from McLeod \& Rieke (1995\markcite{mcl95}),
and a comparison with the result
from the photometry showed good agreement.
The calibration was transferred to the spectra by extracting fluxes with a
rectangular aperture that matched the slit size for the
corresponding spectrum. To obtain an upper limit to
the uncertainties imposed by field rotation at the MMT,
we placed this pseudo slit in 4 orientations
rotated by 45 degrees on the nucleus of each galaxy.
In most cases, the errors due to slit rotation were
the dominant source of uncertainty. These uncertainties and the
statistical ones were propagated through our CO band
correction procedure to obtain a final error on the CO
correction.

\section{Analysis}

\subsection{The CO Band Strength and Circumnuclear Starbursts}

\subsubsection{CO Band Strength and Starbursts}

The goal of this study is to probe the presence of
circumnuclear starbursts in the Seyfert galaxies by
measuring their 2.3$\mu$m first overtone CO bandhead strength.
The strength of this band depends on stellar temperature,
metallicity, and surface gravity/luminosity. These multiple
dependencies suggest that interpretation of the band
would be difficult. However, in the case of nuclear starbursts in
luminous galaxies, the behavior is dominated by the stellar
luminosity.

The best evidence for this conclusion is empirical. From the
work of Ridgway, Wynn-Williams \& Becklin (1994\markcite{rid94}),
we compute that 23 "normal" galaxies recalibrated to
the spectroscopic scale have a CO index of 0.20.
Because of the tight distribution, this value
is very well determined. In comparison, from this same
work we find that the average CO band absorption
index in 14 nuclear starbursts is more widely distributed,
but with an average value of 0.27 (as suggested by Ridgway et al.,
we include only starbursts with $\rm H-K < 0.9).$
The study of luminous infrared galaxies by Goldader et al.
(1995\markcite{gol95}) includes 31 examples dominated by star formation,
for which the average CO index corrected to the spectroscopic
scale is also 0.27. The average photometric CO index of the higher
quality spectra of Engelbracht (1997\markcite{eng97}) is $0.30\pm0.07$.
The data of Oliva et al. (1995\markcite{oli95}) show a similar behavior.

For four low metallicity starbursting galaxies with well developed
CO bands, the band strengths fall close to the average for normal
galaxies (Vanzi \& Rieke, 1997\markcite{van97}).
The metallicities of these galaxies are typically 3 to 5 times lower
than solar. Thus, metallicity does have the expected effect on starburst
CO band strengths, but it is only a modest one. Because of this modest
dependence, it is unlikely that the strong CO
bands in starbursts could arise from high metallicity,
particularly since there is little evidence for higher
than solar metallicities in starbursts.

The starburst CO band strengths measured by Ridgway, Winn-Williams
\& Becklin (1994\markcite{rid94}) and Goldader et al.
(1995\markcite{gol95}) do show a range that
overlaps with the average value for
normal galaxies, so it would be hazardous to conclude that
any given galaxy with a normal CO band depth could not
harbor a starburst. However, there are few if any normal
galaxies with bands as deep as the average for
the starbursts, so deep bands are a relatively reliable
indication of a recent starburst. This conclusion becomes
even stronger in comparing samples of galaxies, since
a few outliers in the distributions will have little
effect on average values of the CO index.

In principle, the CO bands will be weak if
the starburst is so young that hot stars make a significant
contribution to the $\rm K$-band light. This
situation may underly the few starbursts with weak CO. However,
this phase is short-lived; the red supergiants
start to dominate the near infrared after only a few million years,
and continue to do so, until the burst is 50-100 Myr old.

\subsubsection{Measuring CO Band Strengths in Seyfert Galaxies}

AGNs often have spectral energy distributions peaking in the mid infrared
due to emission by hot dust, and
this component could fill in the $2.3\micron$ absorption and
weaken it. To use our spectra to estimate intrinsic
CO band strengths requires that we correct for this
emission. The hot dust emission rises steeply between
the $\rm K$ and $\rm L$(3.5$\mu$m) bands (e.g., McAlary \& Rieke
1988\markcite{mca88}). In most Sy1s,
where the hot dust dominates the emission in L, the
estimates of hot dust emission at 2.3$\mu$m are
too uncertain to allow an accurate correction to
the CO band strength. In most Sy2s, the excess emission above the
stellar contribution at L is relatively weak, however, and we can
conclude that the dust emission at 2.3$\mu$m is small enough
that even approximate corrections can yield a useful
determination of the intrinsic CO band strength.

\placefigure{fig2}

The distribution of observed CO indices is shown on the top panel
in Figure~\ref{fig2}. We assume that the observed
fluxes at $2.3\micron$ are the sums of similar stellar and non-stellar
contributions. We have developed a procedure to disentangle the two based on
a priori knowledge of their typical infrared colors.

(i) Our first step is to assume that the $\rm K$-band emission is
totally of stellar origin. As we shall see, this assumption
is usually not accurate but by repeating the procedure iteratively
we can correct for the remaining non-stellar contribution.

(ii) Next we estimate the stellar component in $\rm L$-band using the
typical $\rm K-L$ color for stars. The infrared colors of
normal galaxy nuclei are well known and are not a
strong function of the stellar population.
We adopted the typical colors of ellipticals as
determined by Aaronson (1977)\markcite{aar77}: $\rm J-H=0.73$,
$\rm H-K=0.21$,
and by Impey, Wynn-Williams \& Becklin (1986)\markcite{imp86}:
$\rm K-L=0.26$.

(iii) Then we subtract the stellar contribution at $\rm L$ from the observed
flux, and obtain the non-stellar $\rm L$-band flux. This step is
facilitated by the equal diameter of $\rm K$
and $\rm L$ band apertures.

(iv) The next step is to use the typical $\rm K-L$ color for the
non-stellar component to estimate the non-stellar contribution
in the $\rm K$-band. Typical colors of the non-stellar
component were inferred from Seyferts which show no or
very weak CO absorption, hence excluding
galaxies with a significant stellar contribution. We selected eight
galaxies with observed CO index~$\leq0.05$ and obtained an average
nuclear color $(K-L)_{nuc}=2.79\pm0.40$ mag,
as a difference between the $\rm K$ flux measured
within the slit aperture and the $\rm L$-band flux
within the circular aperture. Here we assume that the non-stellar
component is concentrated in a point-source, and is fully encompassed
by both the circular apertures and the slit.

(v) We then subtract a featureless spectrum, normalized to the
non-stellar $\rm K$-band flux, from the flux-normalized
$\rm K$-band spectrum. The resulting spectrum is corrected for
dilution.

The procedure has to be repeated to
correct for the inaccurate initial assumption but it converges after only
one or two iterations.  The correction can be applied in other
ways, such as starting at $\rm J$, or $\rm H$ rather than
$\rm K$-band. The final results converge closely to the similar
values.

An underlying assumption in this approach is that the nuclear
sources in the sample galaxies all have similar $\rm K-L$ colors.
It appears that the near infrared nuclear emission is
characterized approximately by a blackbody at the typical
destruction temperature of dust grains
(Rieke \& Lebofsky, 1981\markcite{rie81};
McAlary \& Rieke, 1988\markcite{mca88};
Alonso-Herrero et al., 1998\markcite{alo98}).
As a test, we combined in various fractions a spectrum
of an elliptical galaxy, representing the dominant old stellar
population, with a diluting black body spectrum, and measured the
colors and the CO index of the composite spectra. Most of our galaxies
are consistent with $\rm T_{Black~Body}=1000-1200K$
(Figure~\ref{fig3}), supporting our assumption for the universal nature
of the diluting spectra in Seyfert galaxies.
In addition, the near infrared continuum appears to be "uncovered"
abruptly at an appropriate viewing angle rather than showing a
succession of galaxy types with differing colors in this continuum
(Ruiz, Rieke, \& Schmidt, 1994\markcite{rui94};
Maiolino, Rieke \& Keller, 1995\markcite{mai95};
Quillen et al., 2000b\markcite{qui00b}).
Therefore, this underlying assumption is plausible, although
differences in nuclear color will limit our approach to galaxies
where the corrections are modest.

\placefigure{fig3}

We also assume that the non-stellar nuclear flux
at $\rm L$-band flux comes from a source smaller than our
slit width; that is, that there is no significant level
of extended featureless emission. Typical scale sizes
for the hot dust source component are $\le$ 10 pc, judged from
variability time scales (Rieke \& Lebofsky, 1979\markcite{rie79};
Glass, 1992\markcite{gla92}; Glass, 1997\markcite{gla97};
Glass, 1998\markcite{gla98}; Quillen et al., 2000a\markcite{qui00a}).
The central engine is not expected
to heat up the dust at 0.5-1 kpc from the center or
further, which is the typical linear size of the slit at the distance
of most of our targets. The dust can also be heated by an extended
starburst in the disk. In that case, the $\rm L$-band nuclear component will
be overestimated, leading to an overcorrection of the CO band, and
deduction of a stronger starburst. This effect would only strengthen
our conclusions if we do not detect excess starburst activity.

We minimized the effect of reddening by starting
our correction procedure from the $\rm K$-band. However,
we can put limits on the reddening by comparing the observed $\rm J-H$ colors
of the Seyferts with those of a typical stellar population.
This comparison is not exact because of both
the likely variations in the reddening within our photometric aperture
and the emission associated with the central engine. However, the derived
$\rm J-H$ color excesses are small, so these potential difficulties
are relatively unimportant. We derive that $\rm K-L$ color
excesses for the stars are only a couple of percent,
smaller than or comparable to the observational uncertainties.
Hence, in our analysis we neglected the effects of reddening in
$\rm K-L$. We will discuss in more detail the colors of some of the
galaxies in the next section.

The procedure described above cannot be applied to all of our targets.
Many Type 1 Seyferts have too strong a non-stellar component. We adopted
the criterion for attempting a correction that the
$\rm L$-to-$\rm K$ aperture flux ratio has to be smaller or equal
to 1.0, equivalent to $\rm K-L \leq 0.91$ mag. The limit was
chosen after an error propagation analyses showed that the CO absorption
in galaxies with larger $\rm L$ band excess cannot be reliably corrected
for the non-stellar contribution. Table~\ref{tbl-5} includes all Seyferts
with corrected CO indices. We excluded from further analysis 2237+07
because of the insufficient signal-to-noise ratio at the long wavelength
edge of the K-band. We also discarded galaxies with corrections larger than
25 \% in flux at $\rm K$ to keep the additional uncertainty added
to the corrected CO indices by the correction procedure itself to $<$ 0.01
mag. This constraints our maximum CO index correction to 0.05 mag. 

\placetable{tbl-5}

For comparison purposes we carried out the same procedure on a sample
of starburst galaxies from Engelbracht (1997)\markcite{eng97}. The
photometric data for these galaxies were collected mostly from the
literature (Table~\ref{tbl-6}). Unlike the Seyferts, starburst
galaxies display significant reddening. For most of the targets
Engelbracht (1997)\markcite{eng97} inferred $\rm A_V=3-10$ mag, from
infrared spectroscopy. We also used the infrared colors to supplement
the reddening estimates. We carried out
the CO correction for the starbursts twice. First, we
neglected the extinction. Second, we dereddened the photometry before
correcting the CO band. The reddening accounts
for part of the $\rm K-L$ color excess leading
to smaller CO corrections. This would strengthen a
conclusion for the absence of strong starbursts in Seyferts.

\placetable{tbl-6}

\subsubsection{Results}

The results of the CO correction procedure starting at $\rm K$, as
described in the previous section, are shown in Table~\ref{tbl-5} for
Seyferts and Table~\ref{tbl-7} for starburst galaxies. The histograms
of the CO indices are plotted in
Figure~\ref{fig2}. The top panel shows the observed CO indices, the
middle panel shows the corrected CO indices neglecting the reddening
($\rm A_V=0$) in starbursts, and the bottom panel shows again the corrected
CO indices but the colors of the starbursts were dereddened prior to the CO
correction. Ticks on the horizontal axis show the bin borders.

The CO absorption strength in luminous ellipticals is
narrowly distributed within the 0.18-0.22 bin (we excluded Abell~770 as a
double nucleus galaxy, and M~87 because of its mildly active nucleus).
See also Ridgway et al. (1994).
As shown in Figure~\ref{fig2}, our corrections
to the CO index in the type 2 Seyfert galaxies usually
recovered the CO indices of an underlying old
stellar population, similar to that of the elliptical galaxies.
We do not find evidence for a significant
contribution from young red stars, except in a few of the
active galaxies.

\placetable{tbl-7}

To determine the significance of this result we ran a Monte-Carlo
simulation. We fitted a
Gaussian to the distribution of the corrected CO
indices of starburst galaxies, with reddening taken into account. Next,
we drew random realizations, and counted the cases that match or have
an excess of shallower CO indices compared with the corrected
distribution of the CO indices of Type 1.8-2 Seyfert galaxies. Based
on $10^5$ realizations, there is a probability of only 0.2\% that the
two distributions are drawn from the same parent distribution.

\subsection{Infrared Colors}

Color-color diagrams ($\rm J-H$ vs. $\rm H-K$ and
$\rm H-K$ vs. $\rm K-L$) from
our sample of Seyfert galaxies are shown in Figure~\ref{fig4}, with
$1\sigma$ errors along both axes. Solid circles represent
Seyfert types 1.8 and later, and open circles represent earlier
Seyferts. Crosses and stars represent the typical colors of elliptical
(Frogel et al., 1978\markcite{fro78}) and starburst
(Engelbracht, 1997\markcite{eng97}) galaxies, respectively. The circle
on the right hand side color-color diagram represents the typical
colors of an elliptical galaxy. Giants lie along the solid lines and
supergiants along the dashed lines. Stellar colors are from Johnson
(1966)\markcite{joh66} and Frogel et al. (1978)\markcite{fro78}.
Although the photometric systems may not be exactly the same,
particularly the $\rm L$-band, they are sufficiently close to our system
for qualitative comparison. The giants span the range between G5 and
M8 on left panel and between G5 and M5 on the right. The supergiants
span the range between B0 and M5 on both panels. Reddening vectors
according to the Rieke \& Lebofsky (1985\markcite{rie85}) reddening law
are also shown.

\placefigure{fig4}

Most of our targets are redder then a typical old, red
giant dominated stellar population as represented
by the elliptical galaxy locus. The
spread of colors cannot be explained just by reddening, as in pure
starburst galaxies which are distributed along a line approximately
parallel to the reddening line. There is an additional component,
showing the non-stellar nuclear contribution. It
is present more strongly in $(K-L)$ (Glass \& Moorwood,
1985;\markcite{gla85} Spinoglio et al., 1995;\markcite{spi95}
Alonso-Herrero, Ward \& Kotilainen, 1996;\markcite{alo96}
Alonso-Herrero et al., 1998\markcite{alo98}).

A number of galaxies show peculiar blue colors. In these cases,
we have searched the literature for additional
measurements, shown in Table~\ref{tbl-8}.
Although comparisons are undermined to
some extend by possible variability, from them we can get
a feeling for the external accuracy of our measurements. In the case
of NGC~1144, we believe the best measurement is
$\rm K-L=0.73$ from Carico et al. (1988)\markcite{car88}.
1058+45 is a faint source and given the observational uncertainties,
we attributed the peculiar $\rm K-L$ color to a statistical
error. Our literature
search for Mkn~270 and NGC~7682 yielded measurements consistent with
ours. The blue colors ($\rm H-K$ and $\rm K-L$) of NGC~7682
can be partially
explained if it has a relatively weak non-stellar component.
This explanation is consistent with the observed CO index of 0.18,
which is close to the typical purely
stellar value. Still, the colors are bluer than expected from a giant
dominated stellar population, suggesting perhaps the presence of blue
young stars of early spectral types. In contrast, Mkn~270 has peculiar
$\rm J-H$ and $\rm K-L$ colors, while $\rm H-K$ is close to
the typical values
of other target galaxies. It has a CO index of 0.11, suggesting some
dilution. The observed colors can be produced by a heavily reddened
$(A_V~\approx~6~mag)$ blue population, even younger than in NGC~7682.

\placetable{tbl-8}

\subsection{Comparison with Other Work}

Oliva et al. (1999\markcite{oli99}) have used the near infrared stellar
absorption bands to determine velocity dispersions and mass to
luminosity ratios, $\rm M/L_H$, for active galaxies. Eleven "obscured"
Seyferts (types 1.8-2) have a distribution of $\rm M/L_H$ that is
consistent with a combination of those for normal spirals and starbursts.
In this comparison, we exclude NGC 1052 and M87 because their nuclear
properties are rather different from the other objects in their sample.
Five of these eleven have sufficiently low $\rm M/L_H$ to suggest recent
circumnuclear starbursts. By comparison, in 14 type 1.8-2 Seyferts, we
find two (Mkn~993 and Mkn~461) with deep CO indicative of starbursts.
Because of the broad distribution of CO band strength in starbursts, we
cannot exclude other members of the sample from this group. However,
from the distributions in Figure 3, we can conclude that such events are
present in only a minority of the galaxies we have observed.

The Oliva et al. (1999\markcite{oli99}) sample favors galaxies with
bright nuclei; our study is less biased in this direction. Nonetheless,
the results are consistent in indicating that the near infrared stellar
luminosities in the majority of type 1.8-2 Seyfert galaxies are NOT
dominated by red supergiants from starbursts.

Oliva et al. show that the $\rm M/L_H$ distribution for eight Seyfert 1
galaxies is indistinguishable from that of normal spirals, Because of
the relatively strong nuclear emission of type 1 Seyfert galaxies, we
have been able to measure CO band strengths for only two, but neither
is strong. It therefore appears that the near infrared luminosities of
type 1 Seyfert nuclei are seldom dominated by starburst-generated
supergiants.

We quantified the role of starbursts in our sample by constructing a
set of composite spectra. We added increasing portions of starburst
spectra to an average spectrum of an elliptical galaxy, and ran the
same Monte-Carlo simulation described above. We took an elliptical
galaxy CO strength distribution and added the distribution typical of
starbursts until there was less than a 5\% probability that the CO
strengths in the synthetic spectra were drawn from the same
distribution as the observed distribution of CO strengths in the
Seyfert galaxies. We excluded Mkn~993 from the fitting because its extremely
strong CO index may partially result from the relatively low 
signal-to-noise ratio of that spectrum.
Increasing the starburst component in the composite spectrum quickly
lowers the probability to draw a matching CO index distribution. The
simulation suggests that typically less than 1/3 of the
$\rm K$-band flux originates from a starburst in our sample
of Type 1.8-2 Seyfert galaxies, at the 95\% confidence level.

Let us assume that the starburst and the old population are represented
by single bursts of ages $10^7$ and $10^9$ years, respectively. Then,
the upper limit on the K luminosity implies that the starburst would
contribute no more than 31\% of the bolometric luminosity, 8\% of the
flux in the V-band and 11\% of the flux in the B-band. That is, the
blue continuum of a typical type 1.8-2 Seyfert galaxy in our sample may
have a weak contribution from young blue stars, but is unlikely to be
dominated by such stars. The starburst masses cannot typically exceed
10$^7$, or in the most extreme cases, 10$^8$ M$_\odot$. The corresponding
limits on the starburst luminosities are similar to or lower than those
of the AGNs in these galaxies.

According to the predictions of the same model,
the CO strength reaches the typical value of an old stellar population
about 50-100 Myr after the beginning of the starburst. Thus,
our technique is not sensitive to very old starbursts. However, after
100 Myr, the starburst absolute $\rm K-band$ flux decreases by 3-4 mag,
and its contribution to the total $\rm K-band$ flux
from the galaxy nucleus would become negligible.

\section{Summary}

We obtained $\rm K$ band spectroscopy of a set of Seyfert
galaxies, to measure their stellar first overtone CO band strengths
at $2.3\micron$. We used $\rm JHKL$ aperture photometry to estimate
the non-stellar contribution at this wavelength. We successfully corrected
the CO band for the dilution for 16 galaxies which were not dominated
by the non-stellar component. The comparison to CO indices measured in
elliptical and purely starbursting galaxies yielded no evidence for strong
starbursts in the majority of Seyfert 1.8-2 galaxies.
This result suggests the nuclear activity in many Seyfert 2
galaxies arises independently of any strong circumnuclear
starbursts. We used an evolutionary starburst model to
place an upper limit of $\sim$ 11\% on the contribution to the
B band light from hot, young stars in a typical type 1.8-2 Seyfert.

\acknowledgments

The research has made use of the NASA/IPAC Extragalactic Database
(NED) which is operated by the Jet Propulsion Laboratory, California
Institute of Technology, under contract with NASA. This work has
been supported by NSF grant AST 95-29190. We thank the anonymous
referee for the useful suggestions which helped to improve the paper.

\clearpage

\begin{deluxetable}{rrlllllcrcrrc}
\tablenum{1}
\tablewidth{0pt}
\tablecaption{Log of spectroscopic observations.\label{tbl-1}}
\tiny
\tablehead{
\multicolumn{1}{r}{NGC}&                    \multicolumn{1}{r}{Mkn}&
\multicolumn{1}{c}{R.A. Dec.}&              \multicolumn{1}{c}{Galaxy}&
\multicolumn{1}{c}{Sy\tablenotemark{a}}&    \multicolumn{1}{c}{$V_{rad}$\tablenotemark{b}}&
\multicolumn{1}{c}{Date}&                   \multicolumn{1}{c}{Site}&
\multicolumn{1}{c}{Int.}&                   \multicolumn{1}{c}{$\lambda_c$\tablenotemark{c}}&
\multicolumn{1}{c}{Grating}&                \multicolumn{1}{c}{S/N\tablenotemark{d}}&
\multicolumn{1}{c}{Slit Size}\nl
\multicolumn{2}{c}{(other ID)\tablenotemark{a}}&\multicolumn{1}{c}{(1950)}&
\multicolumn{1}{c}{Type\tablenotemark{a}}&  \multicolumn{1}{c}{Type}&
\multicolumn{1}{c}{km/s}&                   \multicolumn{1}{c}{}&
\multicolumn{1}{c}{}&                       \multicolumn{1}{c}{Time,s}&
\multicolumn{1}{c}{$\mu$m}&                 \multicolumn{1}{c}{l/mm}&
\multicolumn{1}{c}{}&                       \multicolumn{1}{c}{arcsec\tablenotemark{e}}}
\startdata
\multicolumn{13}{c}{Seyfert Galaxies}\nl
-   &334 &00:00:35 +21:40:54&Pec             &1.8&6582$\pm$2  &95/10/09&2.3&1440&2.25&75 &80 &2.4x4.8\nl
\multicolumn{2}{c}{(0048+29,U524)}
         &00:48:53 +29:07:46&(R')SB(s)b      &1  &10770       &95/10/10&2.3&720 &2.25&75 &30 &2.4x4.8\nl
-   &993 &01:22:43 +31:52:36&SAB0/a          &2  &4658$\pm$6  &97/10/12&MMT&960 &2.33&300&15 &1.2x1.6\nl
-   &573 &01:41:23 +02:05:56&(R)SAB(rs)0+    &2  &5174$\pm$23 &95/10/09&2.3&720 &1.60&75 &30 &2.4x4.8\nl
    &    &                  &                &   &            &95/10/09&2.3&720 &2.25&75 &50 &2.4x4.8\nl
    &    &                  &                &   &            &97/10/12&MMT&2160&2.33&300&30 &1.2x1.6\nl
\multicolumn{2}{c}{(0152+06,U1395)}
         &0152:45 +06:22:02&SA(rs)b         &1.9&5208        &95/10/09&2.3&1440&2.25&75 &25 &2.4x4.8\nl
863 &590 &02:12:00 -00:59:58&SA(s)a:         &1.2&7910$\pm$12 &95/10/09&2.3&720 &2.25&75 &90 &2.4x4.8\nl
1144&-   &02:52:39 -00:23:08&RingB           &2  &8648$\pm$14 &97/10/18&MMT&5520&2.33&300&45 &1.2x1.6\nl
-   &3   &06:09:48 +71:03:11&S0:             &2  &4050$\pm$5  &95/03/13&MMT&4320&2.34&600&40 &1.2x1.6\nl
2273&620 &06:45:38 +60:54:13&SB(r)a:         &2  &1849        &95/03/14&MMT&5760&2.32&600&30 &1.2x2.0\nl
3081&-   &09:57:10 -22:35:10&(R$_1$)SAB(r)0/a&2  &2358$\pm$7  &95/03/14&MMT&1920&2.34&600&20 &1.2x1.6\nl
3080&1243&09:57:14 +13:17:03&Sa              &1  &10608       &96/02/06&MMT&1920&2.25&75 &35 &1.2x2.0\nl
3227&-   &10:20:47 +20:07:06&SAB(s) pec      &1.5&1157$\pm$3  &96/02/07&MMT&480 &2.20&75 &70 &1.2x2.4\nl
    &    &                  &                &   &            &94/03/23&2.3&960 &2.31&600&25 &2.4x6.0\nl
    &    &                  &                &   &            &95/03/13&MMT&1440&2.32&600&40 &1.2x2.0\nl
\multicolumn{2}{c}{(1058+45,U6100)}
         &10:58:43 +45:55:22&Sa?             &2  &8778        &96/04/06&MMT&480 &2.20&75 &15 &1.2x2.0\nl
3516&-   &11:03:23 +72:50:20&(R)SB(s)00:     &1.5&2649$\pm$7  &93/11/25&MMT&2880&2.34&600&50 &1.2x1.6\nl
3786&744 &11:37:05 +32:11:11&SAB(rs)a pec    &1.8&2678$\pm$6  &96/02/05&MMT&960 &2.25&75 &80 &1.2x2.0\nl
3982&-   &11:53:52 +55:24:18&SAB(r)b:        &2  &1109$\pm$6  &96/02/10&MMT&1920&2.20&75 &35 &1.2x2.4\nl
4051&-   &12:00:36 +44:48:35&SAB(rs)bc       &1.5&0725$\pm$5  &96/02/10&MMT&1440&2.20&75 &80 &1.2x2.0\nl
    &    &                  &                &   &            &95/03/13&MMT&2880&2.32&600&40 &1.2x1.6\nl
4151&-   &12:08:01 +39:41:02&(R')SAB(rs)ab:  &1.5&0995$\pm$3  &96/02/10&MMT&1440&2.20&75 &140&1.2x2.0\nl
4235&-   &12:14:37 +07:28:09&SA(s)a          &1  &2410$\pm$10 &96/02/10&MMT&1440&2.20&75 &60 &1.2x2.0\nl
    &    &                  &                &   &            &95/03/14&MMT&12960&2.34&600&30&1.2x1.6\nl
4253&766 &12:15:56 +30:05:26&(R')SB(s)a:     &1.5&3876$\pm$16 &96/02/10&MMT&960 &2.20&75 &70 &1.2x2.0\nl
4258&-   &12:16:29 +47:34:53&SAB(s)bc        &1.9&0448$\pm$3  &95/03/13&MMT&1440&2.31&600&40 &1.2x1.6\nl
4388&-   &12:23:15 +12:56:17&SA(s)b: sp      &2  &2524$\pm$1  &96/02/06&MMT&480 &2.25&75 &30 &1.2x2.4\nl
-   &231 &12:54:05 +57:08:38&SA(rs)c? pec    &1  &12651$\pm$6 &94/04/02&1.55&360 &1.60&75 &110&3.6x7.2\nl
    &    &                  &                &   &            &94/04/02&1.55&1080&2.21&75 &90 &3.6x7.2\nl
5033&-   &13:11:09 +36:51:31&SA(s)c          &1.9&0875$\pm$1  &96/02/06&MMT&240 &2.25&75 &60 &1.2x2.4\nl
    &    &                  &                &   &            &95/03/13&MMT&2880&2.32&600&30 &1.2x1.6\nl
-   &789 &13:29:55 +11:21:44&-               &1  &9476$\pm$22 &96/03/29&MMT&2400&2.20&75 &30 &1.2x2.0\nl
\multicolumn{2}{c}{(1335+39,U8621)}
         &13:35:28 +39:24:31&S?              &1.8&6023        &96/03/29&MMT&2880&2.20&75 &25 &1.2x2.0\nl
5252&-   &13:35:44 +04:47:47&S0              &1.9&6926        &96/03/29&MMT&1440&2.20&75 &45 &1.2x2.0\nl
5256&266 &13:36:15 +48:31:48&Compact pec     &2  &8360$\pm$81 &96/03/27&MMT&1800&2.20&75 &35 &1.2x1.6\nl
    &    &                  &                &   &            &96/03/28&MMT&1440&2.20&75 &40 &1.2x2.0\nl
5283&270 &13:39:41 +67:55:28&S0?             &2  &2700$\pm$77 &96/03/29&MMT&960 &2.20&75 &70 &1.2x2.0\nl
    &    &                  &                &   &            &95/03/13&MMT&5280&2.31&600&30 &1.2x1.6\nl
5273&-   &13:39:55 +35:54:21&SA(s)00         &1.9&1089        &96/02/09&MMT&1440&2.20&75 &50 &1.2x2.4\nl
-   &461 &13:45:04 +34:23:52&S               &2  &4856$\pm$32 &96/03/29&MMT&2880&2.20&75 &40 &1.2x2.0\nl
-   &279 &13:51:54 +69:33:13&S0              &1.5&8814$\pm$90 &96/02/09&MMT&480 &2.20&75 &75 &1.2x2.0\nl
5427&-   &14:00:48 -05:47:19&SA(s)c pec      &2  &2730$\pm$4  &95/03/14&MMT&960 &2.33&600&5  &1.2x1.6\nl
5548&1509&14:15:44 +25:22:01&(R')SA(s)0/a    &1.5&5149$\pm$7  &96/02/07&MMT&960 &2.20&75 &80 &1.2x1.6\nl
5674&-   &14:31:22 +05:40:38&SABc            &1.9&7442        &96/03/28&MMT&3360&2.20&75 &65 &1.2x2.4\nl
    &    &                  &                &   &            &95/03/14&MMT&3840&2.37&600&20 &1.2x1.6\nl
5695&686 &14:35:20 +36:47:02&SBb             &2  &4225$\pm$9  &96/03/29&MMT&1440&2.20&75 &40 &1.2x2.0\nl
    &    &                  &                &   &            &95/03/14&MMT&1920&2.34&600&20 &1.2x1.6\nl
-   &841 &15:01:36 +10:37:56&E               &1.5&10860$\pm$90&96/03/29&MMT&480 &2.20&75 &70 &1.2x2.0\nl
5929&-   &15:24:19 +41:50:41&Sab: pec        &2  &2492$\pm$8  &96/03/29&MMT&960 &2.20&75 &45 &1.2x2.0\nl
5940&1511&15:28:51 +07:37:38&SBab            &1  &10122       &96/02/06&MMT&1440&2.20&75 &25 &1.2x2.4\nl
6104&(1614+35)&16:14:40 +35:49:50&-          &1.5&8382$\pm$50 &95/10/10&2.3&2880&2.25&75 &30 &2.4x4.8\nl
(IIZw136)&1513&21:30:01 +09:55:01&-          &1  &18630$\pm$90&95/10/10&2.3&4320&2.25&75 &100&2.4x4.8\nl
\multicolumn{2}{c}{(2237+07,U12138)}
         &22:37:46 +07:47:34&SBa             &1.8&7375$\pm$90 &95/10/07&2.3&1440&2.20&75 &70 &2.4x6.0\nl
    &    &                  &                &   &            &95/10/10&2.3&720 &2.25&75 &50 &2.4x4.8\nl
7469&1514&23:00:44 +08:36:16&(R')SAB(rs)a    &1.2&4892$\pm$2  &94/11/19&MMT&720 &2.35&600&40 &1.2x1.6\nl
7603&530 &23:16:23 -00:01:47&SA(rs)b: pec    &1.5&8851$\pm$22 &95/10/07&2.3&720 &2.20&75 &60 &2.4x6.0\nl
    &    &                  &                &   &            &95/10/10&2.3&720 &2.25&75 &90 &2.4x4.8\nl
7674&533 &23:25:24 +08:30:13&SA(r)bc pec     &2  &8713$\pm$10 &95/10/10&2.3&720 &2.25&75 &70 &2.4x4.8\nl
7682&-   &23:26:31 +03:15:28&SB(r)ab         &2  &5107        &95/10/09&2.3&1440&1.60&75 &30 &2.4x4.8\nl
    &    &                  &                &   &            &95/10/09&2.3&2160&2.25&75 &30 &2.4x4.8\nl
\tableline
\multicolumn{13}{c}{Template Elliptical Galaxies}\nl
\multicolumn{2}{c}{(Abell770A)}
         &09:14:12 +60:38:00&E               &-  &-           &94/04/04&1.55&1860&2.21&75 &90 &3.6x7.2\nl
\multicolumn{2}{c}{(Abell770B)}
         &09:14:12 +60:38:00&E               &-  &-           &94/04/04&1.55&1860&2.21&75 &50 &3.6x5.4\nl
3379&(M105)&10:45:11 +12:50:48&E1            &-  &0920$\pm$10 &94/04/02&1.55&5520&2.21&75 &90 &3.6x7.2\nl
4472&(M49)&12:27:14 +08:16:36&E2/S0          &-  &0868$\pm$8  &94/04/02&1.55&4860&2.21&75 &70 &3.6x7.2\nl
4486&(M87)&12:28:18 +12:39:58&E+0-1 pec      &Sy &1282$\pm$9  &94/04/04&1.55&3060&2.21&75 &110&3.6x7.2\nl
\tablenotetext{a} {Uxxxx = UGCxxxx}
\tablenotetext{b}{From N.E.D.}
\tablenotetext{c}{Central wavelength.}
\tablenotetext{d}{Signal-to-nise ratio per pixel; averaged over the entire spectral range.}
\tablenotetext{e}{Anguar size of the spectral aperture, i.e. the width of the slit x the width of the extracted one-dimensional spectrum.}
\enddata
\end{deluxetable}

\clearpage

\begin{deluxetable}{lrrrrr}
\tablenum{2}
\tablewidth{0pt}
\tablecaption{Equivalent widths for H-band spectral features $W_\lambda(\AA)$.\label{tbl-2}}
\tablehead{
\multicolumn{1}{c}{Galaxy}&     \multicolumn{1}{c}{$MgI$}&
\multicolumn{1}{c}{$FeI$}&      \multicolumn{1}{c}{$SiI$}&
\multicolumn{1}{c}{$\rm CO$}\nl
\multicolumn{1}{c}{Name}&       \multicolumn{1}{c}{$1.504\tablenotemark{a}$}&
\multicolumn{1}{c}{$1.583$}&      \multicolumn{1}{c}{$1.590$}&
\multicolumn{1}{c}{$1.620$}\nl}
\startdata
Mkn573  & $4.2\pm0.3$ & $2.1\pm0.2$ & $3.4\pm0.2$ & $6.2\pm0.4$\nl
Mkn231  & $1.1\pm0.1$ & $0.0\pm0.2$ & $0.7\pm0.1$ & $0.9\pm0.2$\nl
NGC7682 & $4.6\pm0.3$ & $0.5\pm0.3$ & $3.3\pm0.3$ & $9.0\pm1.0$\nl
\tablenotetext{a}{Central wavelengths in $\rm \mu m$.}
\tablecomments{Equivalent line widths are given in $\AA$.}
\enddata
\end{deluxetable}

\clearpage

\begin{deluxetable}{lrrrrr}
\tablenum{3}
\tablewidth{0pt}
\tablecaption{Equivalent line widths $W_\lambda(\AA)$, and CO index for
$\rm K$-band absorption features, for the sample galaxies.\label{tbl-3}}
\tiny
\tablehead{
\multicolumn{1}{c}{Galaxy}&
\multicolumn{1}{c}{$MgI$}&
\multicolumn{1}{c}{$NaI$}&
\multicolumn{1}{c}{$CaI$}&
\multicolumn{1}{c}{MgI}&
\multicolumn{1}{c}{$\rm CO$}\nl
\multicolumn{1}{c}{Name}&
\multicolumn{1}{c}{$2.107\tablenotemark{a}$}&
\multicolumn{1}{c}{$2.207$}&
\multicolumn{1}{c}{$2.264$}&
\multicolumn{1}{c}{$2.281$}&
\multicolumn{1}{c}{$2.230$}}
\startdata
\multicolumn{6}{c}{Seyfert Galaxies}\nl
Mkn334  &$0.6\pm0.1$&$3.0\pm0.4$& $2.4\pm0.2$&$0.0\pm0.1$&0.07\nl
0048+29 &           &$3.4\pm0.2$& $4.5\pm0.4$&           &0.18\nl
Mkn993  &           &$7.3\pm0.6$& $3.9\pm0.7$&$1.9\pm0.3$&0.26\nl
Mkn573  &$0.7\pm0.3$&$3.1\pm0.3$& $1.3\pm0.4$&$1.0\pm0.3$&0.15\nl
0152+06 &           &$3.3\pm0.4$& $3.4\pm0.3$&           &0.16\nl
Mkn590  &$0.7\pm0.3$&$2.4\pm0.1$& $2.2\pm0.2$&$0.0\pm0.1$&0.06\nl
NGC1144 &           &$4.0\pm0.3$& $3.2\pm0.1$&$0.0\pm0.1$&0.15\nl
Mkn3    &           &           &            &$0.3\pm0.1$&0.20\nl
NGC2273 &           &           &            &$0.8\pm0.1$&0.13\nl
NGC3081 &           &           &            &           &0.20\nl
Mkn1243 &$1.5\pm0.5$&$3.5\pm0.2$& $6.7\pm0.2$&$1.5\pm0.2$&0.04\nl
NGC3227 &$0.0\pm0.1$&$3.2\pm0.3$& $0.7\pm0.2$&$0.0\pm0.1$&0.10\nl
1058+45 &$0.3\pm0.1$&$4.6\pm0.4$&$10.0\pm5.0$&$0.0\pm0.2$&0.22\nl
NGC3516 &           &           &            &           &0.06\nl
Mkn744  &$0.5\pm0.2$&$3.2\pm0.2$& $0.7\pm0.2$&$0.0\pm0.1$&0.09\nl
NGC3982 &$0.0\pm0.1$&$6.4\pm0.6$& $5.3\pm0.2$&$0.0\pm0.2$&0.17\nl
NGC4051 &$0.5\pm0.2$&$1.9\pm0.2$& $2.8\pm0.3$&$0.0\pm0.1$&0.06\nl
NGC4151 &$0.3\pm0.1$&$0.9\pm0.1$& $0.4\pm0.2$&$0.0\pm0.1$&0.01\nl
NGC4235 &$0.9\pm0.4$&$3.3\pm0.3$& $2.4\pm0.2$&$0.8\pm0.2$&0.09\nl
Mkn766  &$0.0\pm0.1$&$0.9\pm0.1$& $1.9\pm0.2$&$0.0\pm0.4$&0.02\nl
NGC4258 &           &           &            &$0.4\pm0.1$&0.18\nl
NGC4388 &$1.6\pm0.3$&$2.3\pm0.2$& $5.8\pm0.7$&$0.0\pm0.2$&0.09\nl
Mkn231  &$0.0\pm0.2$&$0.0\pm0.1$& $3.9\pm0.3$&$0.0\pm0.2$&0.01\nl
NGC5033 &$0.0\pm0.2$&$4.4\pm0.3$& $0.7\pm0.1$&$0.9\pm0.2$&0.14\nl
Mkn789  &$0.0\pm0.5$&$4.1\pm0.3$&$12.6\pm1.0$&$0.0\pm0.5$&0.13\nl
1335+39 &$0.0\pm1.0$&$4.7\pm0.3$& $3.7\pm0.6$&$0.0\pm0.5$&0.04\nl
NGC5252 &$0.9\pm0.5$&$2.0\pm0.4$& $3.3\pm1.0$&$1.0\pm0.5$&0.10\nl
Mkn266  &$0.6\pm0.6$&$5.3\pm0.4$&$15.3\pm1.5$&$1.0\pm0.3$&0.17\nl
Mkn270  &$0.6\pm0.4$&$5.2\pm0.3$& $2.3\pm0.5$&$0.8\pm0.2$&0.11\nl
NGC5273 &$0.9\pm0.1$&$5.0\pm0.4$& $4.8\pm0.8$&$2.0\pm0.4$&0.25\nl
Mkn461  &$0.5\pm0.3$&$5.6\pm0.4$& $0.0\pm0.2$&$2.3\pm0.7$&0.19\nl
Mkn279  &$0.0\pm0.3$&$0.3\pm0.2$& $0.8\pm0.2$&$0.6\pm0.2$&0.03\nl
NGC5427 &           &           &            &           &0.05\nl
NGC5548 &$0.4\pm0.3$&$2.7\pm0.1$& $0.0\pm0.1$&$0.0\pm0.1$&0.03\nl
NGC5674 &$0.0\pm0.3$&$4.8\pm0.2$& $4.7\pm0.3$&$0.0\pm0.2$&0.14\nl
Mkn686  &$1.1\pm0.5$&$3.4\pm0.1$& $5.8\pm0.2$&$0.0\pm0.2$&0.16\nl
Mkn841  &$0.9\pm0.1$&$2.1\pm0.2$& $0.0\pm0.2$&$0.0\pm0.2$&0.02\nl
NGC5929 &$0.5\pm0.2$&$6.1\pm0.4$& $3.2\pm0.3$&$0.0\pm0.2$&0.17\nl
NGC5940 &$0.0\pm0.3$&$0.0\pm0.3$& $3.1\pm0.6$&$0.0\pm0.2$&0.03\nl
1614+35 &$0.0\pm0.5$&$5.4\pm0.5$& $5.7\pm0.8$&$0.0\pm0.3$&0.05\nl
IIZw136 &$0.0\pm0.1$&$0.4\pm0.2$& $0.0\pm0.2$&$0.0\pm0.1$&0.01\nl
2237+07 &$0.4\pm0.3$&$3.6\pm0.3$& $2.3\pm0.4$&$1.9\pm0.5$&0.06\nl
NGC7469 &           &           &            &           &0.25\nl
Mkn530  &$0.8\pm0.1$&$3.4\pm0.5$& $0.9\pm0.4$&$0.0\pm0.1$&0.06\nl
Mkn533  &$0.0\pm0.3$&$1.0\pm0.1$& $0.9\pm0.2$&$0.0\pm0.2$&0.03\nl
NGC7682 &$2.1\pm0.2$&$6.8\pm0.3$& $4.5\pm0.3$&$1.0\pm0.3$&0.18\nl
\multicolumn{6}{c}{Template Elliptical Galaxies}\nl
A770A   &$0.9\pm0.1$&$6.9\pm0.3$& $1.7\pm0.3$&$0.0\pm0.2$&0.16\nl
A770B   &$0.0\pm0.2$&$6.7\pm0.3$&$14.1\pm0.5$&$3.3\pm0.5$&0.23\nl
NGC3379 &$0.8\pm0.1$&$5.7\pm0.3$& $3.2\pm0.2$&$0.9\pm0.2$&0.20\nl
NGC4472 &$0.9\pm0.1$&$5.7\pm0.3$& $2.7\pm0.3$&$0.8\pm0.2$&0.20\nl
M87     &$0.9\pm0.1$&$4.2\pm0.2$& $2.4\pm0.1$&$1.0\pm0.2$&0.16\nl
\tablenotetext{a}{Central wavelengths in $\rm \mu m$.}
\tablecomments{Equivalent line widths are given in $\AA$.
Some galaxies have been observed a few times, and the unweighted 
averages from different spectra are given. The second number in 
each column is $1\sigma$ uncertainty except for the CO index, 
where the typical uncertainty for the entire sample is 0.02 (sse 
section 2.3). Missing equivalent line widths indicate lack of 
spectral coverage.}
\enddata
\end{deluxetable}

\clearpage

\begin{deluxetable}{lcrrrrrrrrl}
\tablenum{4}
\tablewidth{0pt}
\tablecaption{Infrared aperture photometry.\label{tbl-4}}
\tiny
\tablehead{
\multicolumn{1}{c}{Galaxy}&                       \multicolumn{1}{c}{Ap.\tablenotemark{a}}&
\multicolumn{1}{c}{$F_J^{ap}$}&                   \multicolumn{1}{c}{$F_H^{ap}$}&
\multicolumn{1}{c}{$F_{K}^{ap}$}&               \multicolumn{1}{c}{$F_L^{ap}$}&
\multicolumn{1}{c}{$F_{K}^{sl}$}&               \multicolumn{1}{c}{$(J-H)^{ap}$}&
\multicolumn{1}{c}{$(H-K)^{ap}$}&               \multicolumn{1}{c}{$(K-L)^{ap}$}&
\multicolumn{1}{c}{Note\tablenotemark{b}}\nl
\multicolumn{1}{c}{Name}&                     \multicolumn{1}{c}{$\arcsec$}&
\multicolumn{1}{c}{mJy}&                      \multicolumn{1}{c}{mJy}&
\multicolumn{1}{c}{mJy}&                      \multicolumn{1}{c}{mJy}&
\multicolumn{1}{c}{mJy}&                      \multicolumn{1}{c}{mag}&
\multicolumn{1}{c}{mag}&                      \multicolumn{1}{c}{mag}&
\multicolumn{1}{c}{}}
\startdata
 Mkn334&8.6&$13.5\pm0.7$&$ 21.1\pm1.1$&$ 29.0\pm1.5$&$ 41.0\pm2.0$&$21.5\pm1.5$&$0.92\pm0.08$&$0.85\pm0.08$&$ 1.30\pm0.08$&\nl
0048+29&5.4&$10.4\pm0.5$&$ 13.9\pm0.7$&$ 13.1\pm0.7$&$ 10.0\pm1.0$&$ 7.9\pm0.7$&$0.75\pm0.08$&$0.44\pm0.08$&$ 0.63\pm0.12$&\nl
 Mkn993&8.6&$13.8\pm0.7$&$ 19.0\pm0.9$&$ 16.0\pm0.8$&$ 10.0\pm1.0$&$ 2.3\pm0.8$&$0.78\pm0.08$&$0.32\pm0.08$&$ 0.42\pm0.12$&\nl
 Mkn573&8.6&$22.0\pm1.1$&$ 29.5\pm1.5$&$ 26.3\pm1.3$&$ 23.0\pm1.1$&$14.1\pm1.3$&$0.75\pm0.08$&$0.38\pm0.08$&$ 0.78\pm0.08$&$2.4\times4.8$\nl
       &   &            &             &             &             &$ 4.9\pm1.3$&             &             &              &$1.2\times1.6$\nl
0152+06&5.4&$ 5.2\pm0.5$&$  7.0\pm0.7$&$  8.0\pm0.8$&$  6.0\pm0.6$&$ 5.4\pm0.8$&$0.76\pm0.15$&$0.65\pm0.15$&$ 0.62\pm0.15$&\nl
 Mkn590&5.4&$20.0\pm1.0$&$ 25.0\pm1.2$&$ 40.0\pm2.0$&$ 64.0\pm3.2$&$29.9\pm2.0$&$0.68\pm0.08$&$1.01\pm0.08$&$ 1.44\pm0.08$&\nl
NGC1144&5.4&$12.0\pm0.6$&$ 16.5\pm0.8$&$ 14.6\pm0.7$&$  7.0\pm0.7$&$ 1.8\pm0.7$&$0.78\pm0.08$&$0.37\pm0.08$&$ 0.13\pm0.12$&\nl
Mkn1243&5.4&$ 5.5\pm0.6$&$  7.9\pm0.8$&$  9.5\pm0.9$&$ 14.5\pm0.7$&$ 2.3\pm1.0$&$0.83\pm0.15$&$0.70\pm0.15$&$ 1.39\pm0.12$&\nl
NGC3227&8.6&$63.0\pm3.2$&$ 78.0\pm3.9$&$ 78.0\pm3.9$&$ 71.0\pm3.5$&$16.3\pm3.9$&$0.67\pm0.08$&$0.50\pm0.08$&$ 0.83\pm0.08$&$1.2\times2.0$\nl
       &   &            &             &             &             &$12.9\pm3.9$&             &             &              &$1.2\times1.6$\nl
       &   &            &             &             &             &$45.9\pm3.9$&             &             &              &$2.4\times6.0$\nl
NGC3362&5.4&$ 4.2\pm0.4$&$  5.4\pm0.5$&$  4.5\pm0.5$&$  4.0\pm1.0$&            &$0.71\pm0.15$&$0.32\pm0.15$&$ 0.78\pm0.16$&\nl
1058+45&5.4&$ 7.2\pm0.7$&$  9.7\pm1.0$&$  8.2\pm0.8$&$  3.3\pm0.3$&$ 2.2\pm0.8$&$0.76\pm0.15$&$0.32\pm0.15$&$-0.06\pm0.15$&\nl
 Mkn744&8.6&$19.8\pm1.0$&$ 25.9\pm1.3$&$ 27.8\pm1.4$&$ 31.0\pm1.5$&$ 8.4\pm1.4$&$0.73\pm0.08$&$0.58\pm0.08$&$ 1.05\pm0.08$&\nl
NGC3982&5.4&$11.2\pm0.6$&$ 12.2\pm0.6$&$ 10.2\pm0.5$&$  7.1\pm0.9$&            &$0.53\pm0.08$&$0.32\pm0.08$&$ 0.52\pm0.12$&\nl
NGC4051&8.6&$40.0\pm2.0$&$ 49.0\pm2.5$&$ 65.0\pm3.2$&$ 77.0\pm3.8$&$14.9\pm3.3$&$0.65\pm0.08$&$0.81\pm0.08$&$ 1.11\pm0.08$&$1.2\times1.6$\nl
       &   &            &             &             &             &$17.7\pm3.3$&             &             &              &$1.2\times2.0$\nl
NGC4151&8.6&$105.0\pm5.2$&$130.0\pm6.5$&$194.0\pm9.7$&$ 344\pm17.2$&$35.3\pm9.8$&$0.67\pm0.08$&$0.94\pm0.08$&$ 1.55\pm0.08$&\nl
NGC4235&5.4&$23.0\pm1.1$&$ 35.4\pm1.8$&$ 29.9\pm1.5$&$ 22.0\pm1.1$&$ 4.8\pm1.5$&$0.90\pm0.08$&$0.32\pm0.08$&$ 0.59\pm0.08$&$1.2\times1.6$\nl
       &   &            &             &             &             &$ 5.7\pm1.5$&             &             &              &$1.2\times2.0$\nl
 Mkn766&5.4&$16.3\pm0.8$&$ 26.7\pm1.3$&$ 43.2\pm2.2$&$ 86.0\pm4.3$&$11.8\pm2.2$&$0.97\pm0.08$&$1.02\pm0.08$&$ 1.68\pm0.08$&\nl
 Mkn231&5.8&$43.0\pm2.2$&$ 77.0\pm3.8$&$175.0\pm8.8$&$   360\pm18$&$73.5\pm8.8$&$1.07\pm0.08$&$1.39\pm0.08$&$ 1.71\pm0.08$&\nl
NGC5033&8.6&$44.1\pm2.2$&$ 61.6\pm3.1$&$ 47.8\pm2.4$&$ 43.0\pm2.2$&$ 6.0\pm2.4$&$0.80\pm0.08$&$0.23\pm0.08$&$ 0.81\pm0.08$&$1.2\times1.6$\nl
       &   &            &             &             &             &$ 7.4\pm2.4$&             &             &              &$1.2\times2.0$\nl
       &   &            &             &             &             &$ 8.1\pm2.4$&             &             &              &$1.2\times2.4$\nl
1335+39&5.4&$ 4.5\pm0.4$&$  6.7\pm0.7$&$  7.2\pm0.7$&$ 10.2\pm0.5$&$ 2.0\pm0.7$&$0.87\pm0.15$&$0.58\pm0.15$&$ 1.31\pm0.12$&\nl
NGC5252&5.4&$10.6\pm0.5$&$ 14.9\pm0.7$&$ 14.9\pm0.7$&$ 19.3\pm1.0$&$ 4.2\pm0.8$&$0.80\pm0.08$&$0.50\pm0.08$&$ 1.21\pm0.08$&\nl
 Mkn266&8.6&$ 6.0\pm0.6$&$  8.8\pm0.9$&$  8.7\pm0.9$&$  7.0\pm0.7$&$ 2.6\pm0.9$&$0.85\pm0.15$&$0.49\pm0.15$&$ 0.69\pm0.15$&$1.2\times1.6$\nl
       &   &            &             &             &             &$ 3.3\pm1.0$&             &             &              &$1.2\times2.0$\nl
 Mkn270&5.8&$16.0\pm0.8$&$ 18.0\pm0.9$&$ 17.0\pm0.9$&$  7.0\pm0.7$&$ 5.3\pm0.9$&$0.56\pm0.08$&$0.44\pm0.08$&$-0.04\pm0.12$&$1.2\times2.0$\nl
       &   &            &             &             &             &$ 4.6\pm0.9$&             &             &              &$1.2\times1.6$\nl
 Mkn461&5.4&$ 8.5\pm0.9$&$ 10.8\pm0.5$&$  9.1\pm0.9$&$  8.5\pm0.9$&$ 2.6\pm0.9$&$0.69\pm0.12$&$0.32\pm0.12$&$ 0.85\pm0.15$&\nl
 Mkn279&8.6&$18.0\pm0.9$&$ 24.0\pm1.2$&$ 23.0\pm1.1$&$ 32.0\pm1.6$&$ 7.2\pm1.2$&$0.75\pm0.08$&$0.46\pm0.08$&$ 1.29\pm0.08$&\nl
 IC4397&5.4&$ 4.7\pm0.5$&$  6.5\pm0.7$&$  5.4\pm0.5$&$  5.8\pm1.4$&            &$0.79\pm0.15$&$0.32\pm0.15$&$ 0.99\pm0.16$&\nl
NGC5548&8.6&$27.0\pm1.4$&$ 36.0\pm1.8$&$ 52.0\pm2.6$&$ 91.0\pm4.6$&$12.9\pm2.6$&$0.75\pm0.08$&$0.90\pm0.08$&$ 1.54\pm0.08$&\nl
NGC5674&5.4&$11.1\pm0.6$&$ 15.2\pm0.8$&$ 15.1\pm0.8$&$ 15.5\pm0.8$&$ 4.9\pm0.8$&$0.78\pm0.08$&$0.49\pm0.08$&$ 0.96\pm0.08$&$1.2\times1.6$\nl
       &   &            &             &             &             &$ 5.9\pm0.8$&             &             &              &$1.2\times2.4$\nl
 Mkn686&5.4&$13.3\pm0.7$&$ 17.2\pm0.9$&$ 13.9\pm0.7$&$  8.9\pm0.9$&$ 3.0\pm0.7$&$0.71\pm0.08$&$0.27\pm0.08$&$ 0.44\pm0.12$&$1.2\times1.6$\nl
       &   &            &             &             &             &$ 3.5\pm0.7$&             &             &              &$1.2\times2.0$\nl
NGC5929&5.4&$14.6\pm0.7$&$ 20.2\pm1.0$&$ 17.2\pm0.9$&$ 13.7\pm0.7$&$ 4.1\pm0.9$&$0.79\pm0.08$&$0.33\pm0.08$&$ 0.68\pm0.08$&\nl
NGC5940&5.4&$ 4.7\pm0.5$&$  7.0\pm0.7$&$  8.3\pm0.8$&$ 10.7\pm0.8$&            &$0.87\pm0.15$&$0.70\pm0.15$&$ 1.19\pm0.16$&\nl
1614+35&5.4&$ 3.6\pm0.4$&$  5.6\pm0.6$&$  5.6\pm0.6$&$  6.4\pm1.0$&            &$0.91\pm0.15$&$0.52\pm0.15$&$ 1.06\pm0.16$&\nl
2237+07&8.6&$12.0\pm0.6$&$ 17.8\pm0.9$&$ 19.4\pm1.0$&$ 16.0\pm0.8$&$12.3\pm1.0$&$0.86\pm0.08$&$0.60\pm0.08$&$ 0.72\pm0.08$&$2.4\times6.0$\nl
       &   &            &             &             &             &$11.9\pm1.0$&             &             &              &$2.4\times4.8$\nl
NGC7469&8.6&$57.0\pm2.8$&$ 74.0\pm3.7$&$123.0\pm6.2$&$  163\pm8.1$&$18.5\pm6.2$&$0.72\pm0.08$&$1.05\pm0.08$&$ 1.23\pm0.08$&\nl
 Mkn530&8.6&$17.0\pm0.9$&$ 25.0\pm1.2$&$ 25.0\pm1.2$&$ 25.0\pm1.2$&$16.1\pm1.3$&$0.85\pm0.08$&$0.50\pm0.08$&$ 0.93\pm0.08$&$2.4\times6.0$\nl
       &   &            &             &             &             &$15.4\pm1.3$&             &             &              &$2.4\times4.8$\nl
 Mkn533&8.6&$10.1\pm0.5$&$ 15.6\pm0.8$&$ 24.2\pm1.2$&$ 46.0\pm2.3$&$16.1\pm1.2$&$0.91\pm0.08$&$0.98\pm0.08$&$ 1.63\pm0.08$&\nl
NGC7682&8.6&$10.5\pm0.5$&$ 14.9\pm0.7$&$  9.5\pm0.9$&$  4.6\pm0.5$&$ 5.7\pm1.0$&$0.81\pm0.08$&$0.01\pm0.12$&$ 0.14\pm0.15$&\nl
\tablenotetext{a}{Aperture diameter, arcsec.}
\tablenotetext{b}{Slit size in arcsec, if more then one spectra were taken}
\tablecomments{The table contains: $\rm JHKL$ fluxes within the given aperture diameter, centered at the
nucleus, $\rm K$ flux within the slit, used to calibrate the spectra, and the colors derived from the 
aperture fluxes, using the absolute calibration of Campins, Rieke \& Lebofsky (1985)\markcite{cam85}.
The second number in each column is $1\sigma$ uncertainty.}
\enddata
\end{deluxetable}

\begin{deluxetable}{lccccccccl}
\scriptsize
\tablenum{5}
\tablewidth{0pt}
\tablecaption{Corrected CO indices for Seyfert galaxies.\label{tbl-5}}
\tablehead{
\multicolumn{1}{c}{Name}& \multicolumn{1}{c}{$(L/K)_{ap}$}&
\multicolumn{1}{c}{$\rm CO_{obs}$}&
\multicolumn{1}{c}{$\rm CO^H_{cor}$}    &\multicolumn{1}{c}{$\%^H_{n-st}$}&
\multicolumn{1}{c}{Iter.}&
\multicolumn{1}{c}{$\rm CO^{K}_{cor}$}&\multicolumn{1}{c}{$\%^{K_s}_{n-st}$}&
\multicolumn{1}{c}{Iter.}&
\multicolumn{1}{c}{Note}\nl
\multicolumn{1}{c}{1}& 
\multicolumn{1}{c}{2}&
\multicolumn{1}{c}{3}&
\multicolumn{1}{c}{4}&
\multicolumn{1}{c}{5}&
\multicolumn{1}{c}{6}&
\multicolumn{1}{c}{7}&
\multicolumn{1}{c}{8}&
\multicolumn{1}{c}{9}&
\multicolumn{1}{c}{10}}
\startdata
0048+29&0.76&0.18&0.21&10&1&0.20& 6&2&\nl
Mkn993 &0.62&0.26&0.33&18&1&0.29&10&1&\nl
Mkn573 &0.87&0.15&0.21&29&1&0.19&21&1&\nl
0152+06&0.75&0.16&0.18&11&1&0.18& 5&2&\nl
NGC1144&0.48&0.15&0.15& 2&1&0.15& 0&0&\nl
NGC3227&0.91&0.10&0.15&40&1&0.13&28&1&\nl
1058+45&0.40&0.22&0.22& 0&0&0.22& 0&0&$\rm L-K<(L-K)_{Sy1}$\nl
NGC4235&0.74&0.09&0.11&27&1&0.10&19&1&\nl
NGC5033&0.90&0.14&0.23&48&1&0.22&42&1&\nl
Mkn266 &0.80&0.17&0.22&22&1&0.20&14&1&\nl
Mkn270 &0.41&0.11&0.11& 0&0&0.11& 0&0&$\rm L-K<(L-K)_{Sy1}$\nl
Mkn461 &0.93&0.19&0.26&29&1&0.24&23&1&\nl
Mkn686 &0.64&0.16&0.18&11&1&0.18& 7&1&\nl
NGC5929&0.80&0.17&0.23&25&1&0.21&18&1&\nl
2237+07&0.82&0.06&0.08&14&1&0.08& 8&2&\nl
NGC7682&0.48&0.18&0.18& 0&0&0.18& 0&0&$\rm L-K<(L-K)_{Sy1}$\nl
\tablecomments{Columns 4 and 7 contain the corrected CO indices with 
correction procedure started from $\rm H$ and $\rm K$ respectively. 
Columns 5 and 8 show the non-stellar part of the $\rm K$ flux. The 
number of iterations are presented in columns 6 and 9. Zero iterations 
indicates that the galaxy has $\rm K-L$ bluer then the adopted pure 
non-stellar color $(K-L)_{Sy1}=2.79$, and we assumed negligible 
non-stellar flux at $\rm K$.}
\enddata
\end{deluxetable}

\begin{deluxetable}{lccrrrrrrrc}
\scriptsize
\tablenum{6}
\tablewidth{0pt}
\tablecaption{Photometric data for the control sample of starburst galaxies.\label{tbl-6}}
\tablehead{
\multicolumn{1}{c}{Name}&                   \multicolumn{1}{c}{Slit Size}&
\multicolumn{1}{c}{Eff\tablenotemark{a}.}&  \multicolumn{1}{c}{Phot.}&
\multicolumn{1}{c}{$F^{ap}_J$}&             \multicolumn{1}{c}{$F^{ap}_H$}&  \multicolumn{1}{c}{$F^{ap}_{K}$}&
\multicolumn{1}{c}{$F^{sl}_{K}$}&         \multicolumn{1}{c}{$F^{ap}_L$}&
\multicolumn{1}{c}{$\rm A_V$}&                  \multicolumn{1}{c}{Ref.}\nl
\multicolumn{1}{c}{}&              \multicolumn{1}{c}{$\arcsec$}&
\multicolumn{1}{c}{Ap.,$\arcsec$}& \multicolumn{1}{c}{Ap.,$\arcsec$}&
\multicolumn{1}{c}{mJy}&           \multicolumn{1}{c}{mJy}&       \multicolumn{1}{c}{mJy}&
\multicolumn{1}{c}{mJy}&           \multicolumn{1}{c}{mJy}&
\multicolumn{1}{c}{Mag}&           \multicolumn{1}{c}{}}
\startdata
NGC253 &$2.4\times12.0$&6.1& 6.0& 115.0& 226.7& 280.6& 278.1& 390.3& 7.4& 1,8\tablenotemark{b} \nl
NGC660 & $2.4\times9.6$&5.4& 5.0&  27.9&  40.9&  36.7&  69.2&  34.7& 6.4& 6 \nl
Maffei2& $2.4\times6.0$&4.3&slit&  16.6&  45.6&  55.5&  55.5&   -  & 8.5& 1 \nl
IC342  & $2.4\times6.0$&4.3& 3.8&  49.0&  74.0&  71.0&  87.1&  44.0& 5.1& 9 \nl
NGC1614& $2.4\times6.0$&4.3& 6.0&  32.3&  52.4&  51.5&  27.6&  55.9& 3.5& 11 \nl
NGC2146& $2.4\times6.0$&4.3&17.0& 132.1& 253.2& 270.5&  94.3& 285.7& 6.4& 2 \nl
NGC2782& $2.4\times6.0$&4.3&17.0&  55.1&  73.7&  62.5&  22.1&  65.6& 2.9& 2 \nl
M82    &$2.4\times26.4$&8.9& 7.8& 161.0& 350.0& 475.0&   -  & 436.0&10.2& 10 \nl
NGC3079& $2.4\times6.0$&4.3& 6.0&  29.2&  66.6&  82.4&  34.7&  63.9& 7.8& 4 \nl
NGC3628& $2.4\times6.0$&4.3&slit&  06.2&  14.8&  19.4&  19.4&   -  & 9.7& 1 \nl
NGC4102& $2.4\times6.0$&4.3& 3.8&  80.5& 118.0& 117.0&  87.1&  89.0& 5.7& 5 \nl
NGC4194& $2.4\times6.0$&4.3&13.0&  73.3&  78.6&  67.3&  24.9&  69.0& 3.5& 3 \nl
NGC4339& $2.4\times6.0$&4.3&slit&  15.0&  21.6&  17.2&  17.2&   -  & 1.0& 1 \nl
NGC5990& $2.4\times6.0$&4.3&slit&  16.6&  23.7&  27.6&  27.6&   -  & 7.7& 1 \nl
NGC6000& $2.4\times6.0$&4.3&slit&  30.3&  37.3&  37.3&  37.3&   -  & 5.1& 1 \nl
NGC6240& $2.4\times6.0$&4.3& 9.0&  27.9&  50.5&  53.5&  31.3&  43.2& 9.9& 3 \nl
NGC6946& $2.4\times7.2$&4.7& 3.0&  23.8&  35.3&  36.3&   -  &   -  & 4.3& 1 \nl
NGC7714& $2.4\times6.0$&4.3& 5.0&  20.4&  25.6&  21.5&  18.5&  22.4& 4.3& 4 \nl
Mkn331 & $2.4\times6.0$&4.3& 5.0&  26.0&  36.4&  34.9&  23.8&  29.1& 4.8& 7 \nl
\tablenotetext{a}{The effective aperture is a circle with the same area 
as the slit aperture.}
\tablenotetext{b}{1 for $\rm JHK$, 8 for $\rm K-L$}
\tablecomments{References are for the aperture photometry only.
The flux within the slit is from Engelbracht (1997)\markcite{eng97}, 
who gives $\rm K_s$, instead of $\rm K$, unlike the other sources. 
The difference between the two filters is of order of 0.01 mag, 
insignificant compared to the typical uncertainties of 0.05-0.10 mag.}
\tablerefs{
(1) Engelbracht (1997)\markcite{eng97};
(2) Hunt \& Giovanardi (1992)\markcite{hun92};
(3) Allen (1976)\markcite{ale76};
(4) Lawrence et al. (1985)\markcite{law85};
(5) Roche et al. (1991)\markcite{roc91};
(6) Brindle et al. (1991)\markcite{bri91};
(7) Carico et al. (1988)\markcite{car88};
(8) Rieke \& Low (1975)\markcite{rie75};
(9) Becklin et al. (1980)\markcite{bec80};
(10) Rieke et al. (1980)\markcite{rie80};
(11) Glass \& Moorwood (1985)\markcite{gla85}
}
\enddata
\end{deluxetable}

\begin{deluxetable}{lrrrcrrrc}
\scriptsize
\tablenum{7}
\tablewidth{0pt}
\tablecaption{Observed and corrected CO indices for the control sample of starburst galaxies.\label{tbl-7}}
\tablehead{
\multicolumn{1}{c}{Name}&
\multicolumn{1}{c}{$\rm CO_{obs}$}&
\multicolumn{3}{c}{$\rm A_V=0$}&
\multicolumn{4}{c}{$\rm A_V\neq0$}\nl
\multicolumn{1}{c}{}&
\multicolumn{1}{c}{}&
\multicolumn{1}{c}{$\rm CO^H_{cor}$}&
\multicolumn{1}{c}{$\rm CO^{K}_{cor}$}&
\multicolumn{1}{c}{Iter.}&
\multicolumn{1}{c}{$\rm A_V$}&
\multicolumn{1}{c}{$\rm CO^H_{cor}$}&
\multicolumn{1}{c}{$\rm CO^{K}_{cor}$}&
\multicolumn{1}{c}{Iter.}\nl
\multicolumn{1}{c}{1}&
\multicolumn{1}{c}{2}&
\multicolumn{1}{c}{3}&
\multicolumn{1}{c}{4}&
\multicolumn{1}{c}{5}&
\multicolumn{1}{c}{6}&
\multicolumn{1}{c}{7}&
\multicolumn{1}{c}{8}&
\multicolumn{1}{c}{9}}
\startdata
NGC253 &0.24&0.31&0.29&1& 7.4&0.27&0.26&1\nl
NGC660 &0.22&0.23&0.23&1& 6.4&0.22&0.22&1\nl
Maffei2&0.22&0.22&0.22&0& 8.5&0.22&0.22&0\nl
IC342  &0.26&0.27&0.26&1& 5.1&0.25&0.27&2\nl
NGC1614&0.28&0.28&0.28&0& 3.5&0.28&0.28&0\nl
NGC2146&0.18&0.27&0.24&1& 6.4&0.21&0.20&1\nl
NGC2782&0.14&0.19&0.18&1& 2.9&0.17&0.17&1\nl
M82    &0.27&0.27&0.27&0&10.2&0.27&0.27&0\nl
NGC3079&0.17&0.22&0.19&1& 7.8&0.17&0.17&1\nl
NGC3628&0.24&0.24&0.24&0& 9.7&0.24&0.24&0\nl
NGC4102&0.19&0.21&0.22&1& 5.7&0.19&0.20&2\nl
NGC4194&0.15&0.21&0.20&1& 3.5&0.18&0.18&1\nl
NGC4339&0.25&0.25&0.25&0& 1.0&0.25&0.25&0\nl
NGC5990&0.12&0.12&0.12&0& 7.7&0.12&0.12&1\nl
NGC6000&0.25&0.25&0.25&0& 5.1&0.25&0.25&0\nl
NGC6240&0.18&0.22&0.21&2& 9.9&0.17&0.18&1\nl
NGC6946&0.26&0.26&0.26&0& 4.3&0.26&0.26&0\nl
NGC7714&0.26&0.30&0.29&1& 4.3&0.27&0.28&1\nl
Mkn331 &0.22&0.25&0.26&2& 4.8&0.23&0.23&2\nl
\tablecomments{Columns 1-9 are the same as in Table~\ref{tbl-5}.}
\enddata
\end{deluxetable}

\begin{deluxetable}{lrcccc}
\scriptsize
\tablenum{8}
\tablewidth{0pt}
\tablecaption{Aperture photometry of Seyfert galaxies from 
the literature.\label{tbl-8}}
\tablehead{
\multicolumn{1}{c}{Name}   &\multicolumn{1}{c}{Ap,$\arcsec$}&
\multicolumn{1}{c}{$\rm J-H$}  &\multicolumn{1}{c}{$\rm H-K$}&
\multicolumn{1}{c}{$\rm K-L$}&\multicolumn{1}{c}{Ref.}}
\startdata
NGC1144& 4.0&0.89&0.38&     0.55&1\nl
       & 5.0&0.72&0.44&     0.73&2\nl
       & 5.4&1.01&0.47&         &3\nl
       & 8.5&1.43&0.58&         &4\nl
       &12.0&0.75&0.41&         &5\nl
       &30.0&0.75&0.21&         &6\nl
Mkn270 & 5.9&0.50&0.46&    -0.07&7\nl
       &15.0&0.50&0.35&$\le$0.84&8\nl
NGC7682& 8.5&0.80&0.02&         &4\nl
\tablerefs{
(1) Joy \& Ghigo (1988)\markcite{joy88};
(2) Carico et al. (1988)\markcite{car88};
(3) Bushouse \& Stanford (1985)\markcite{bus92};
(4) Edelson, Malkan \& Rieke (1987)\markcite{ede87};
(5) Cutri \& McAlary (1985)\markcite{cut85};
(6) Spinoglio et al. (1995)\markcite{spi95};
(7) Rieke (1978)\markcite{rie78};
(8) McAlary, McLaren \& Crabtree (1979)\markcite{mca79}}
\enddata
\end{deluxetable}

\clearpage

\figcaption[fig01_xx.eps]{Spectra of the sample galaxies, in order of 
increasing R.A., continuum-divided (i.e. all continua were normalized 
to 1) and shifted vertically for display purposes by adding shiftss 
(from bottom to top: 0, 1, 2, ..., 8 for the K-band spectra, and 0, 
0.8, 1.6 for the H-band), i.e. for each spectrum, the flux interval 
from zero to the continuum level ($F^{sl}_K$, mJy in Table 4) spans one 
arbitrary flux unit. $1\sigma$ noise spectrum (see section2.1) is shown 
above each spectrum, on the same scale. For galaxies with multiple 
spectra all spectra are shown. The galaxy names and Seyfert types are 
indicated. Vertical dot lines show some features of interest. (a)-(d) 
K-band spectra. (e) H-band spectra.
\label{fig1}}

\figcaption[fig02.eps]{Histogram of the observed and corrected values
of the CO index in Type 1-1.5 Seyferts (dotted line), in Type 1.8-2
Seyferts (solid line), and in starbursts (dashed line). The top panel 
shows the observed CO indices, the middle panel shows the CO indices
corrected starting from $\rm K$-band neglecting the reddening ($\rm A_V=0$) 
in starbursts, and the bottom panel shows again the corrected
CO indices but the colors of starburst have been dereddened prior to
the CO correction. Ticks on the horizontal axis show the bin borders.
The ellipticals fall into 0.18-0.22 bin. See section 3.1 for details.
\label{fig2}}

\figcaption[fig04.eps]{CO index vs. $\rm K-L$ diagram. The solid
dots represent Seyfert galaxies of Type 1.8 and later, and the open
circles represent the earlier Type Seyfert galaxies. The grid represents
the CO index and infrared colors of a combination of an elliptical galaxy
spectrum and a diluting black body (solid lines, the temperature is
labeled on the top) spectrum, in various ratios (dashed lines, the black 
body-to-elliptical spectrum ratio is labeled on the right).
\label{fig3}}

\figcaption[fig03.eps]{Color-color diagrams for Seyfert galaxies.
See section 3.2 for explanations.
\label{fig4}}

\clearpage
\plotone{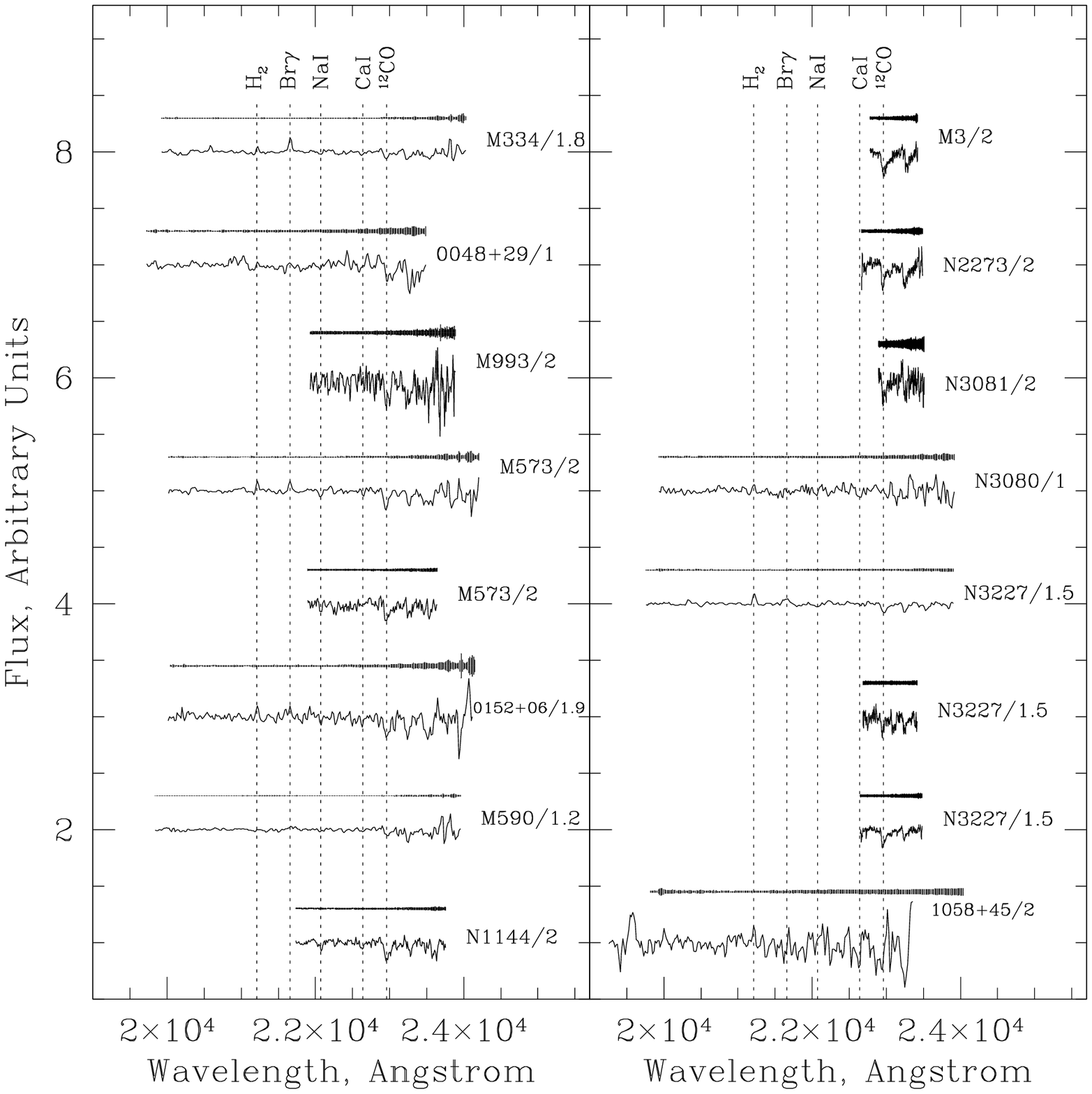}
\clearpage
\plotone{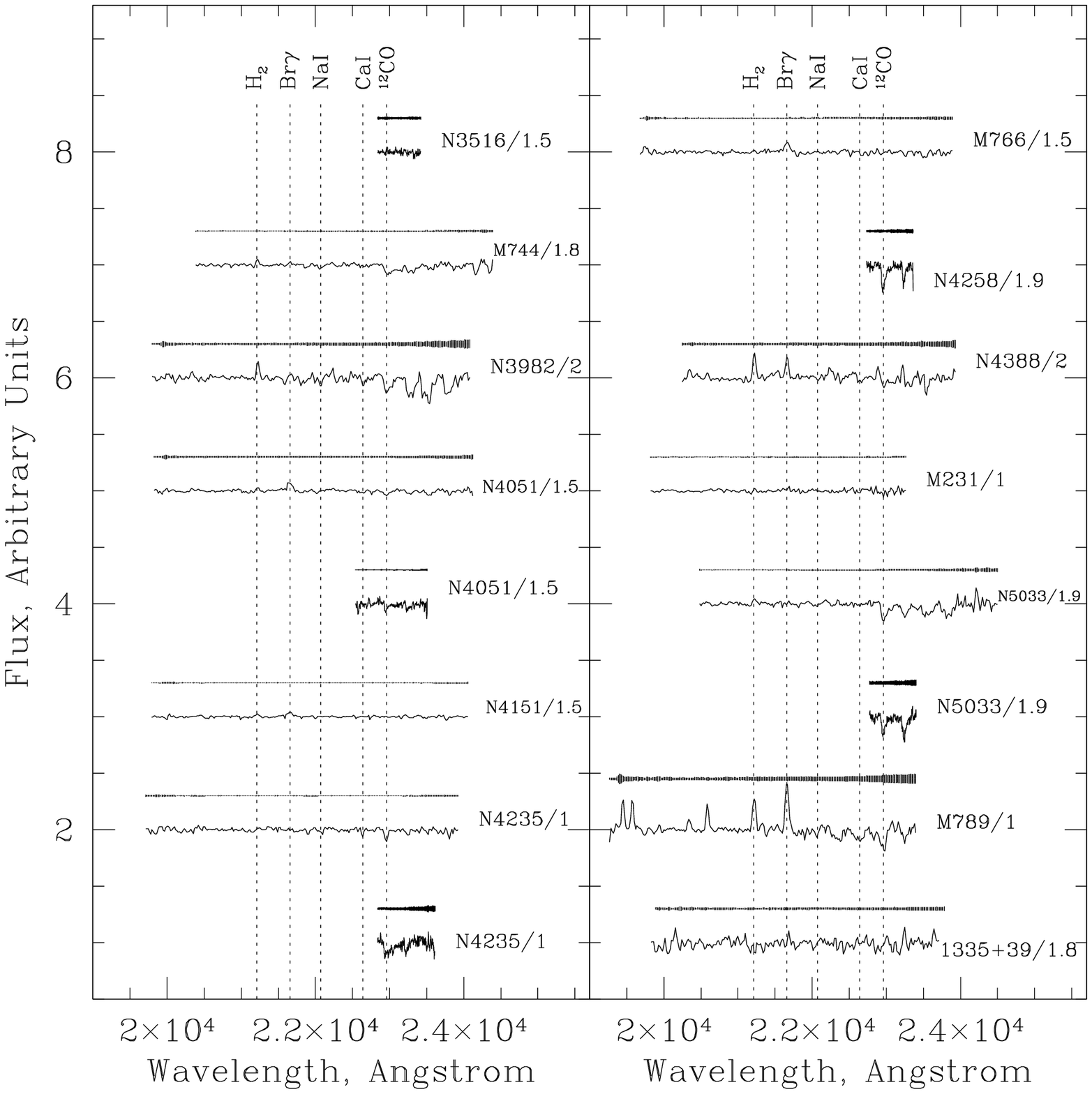}
\clearpage
\plotone{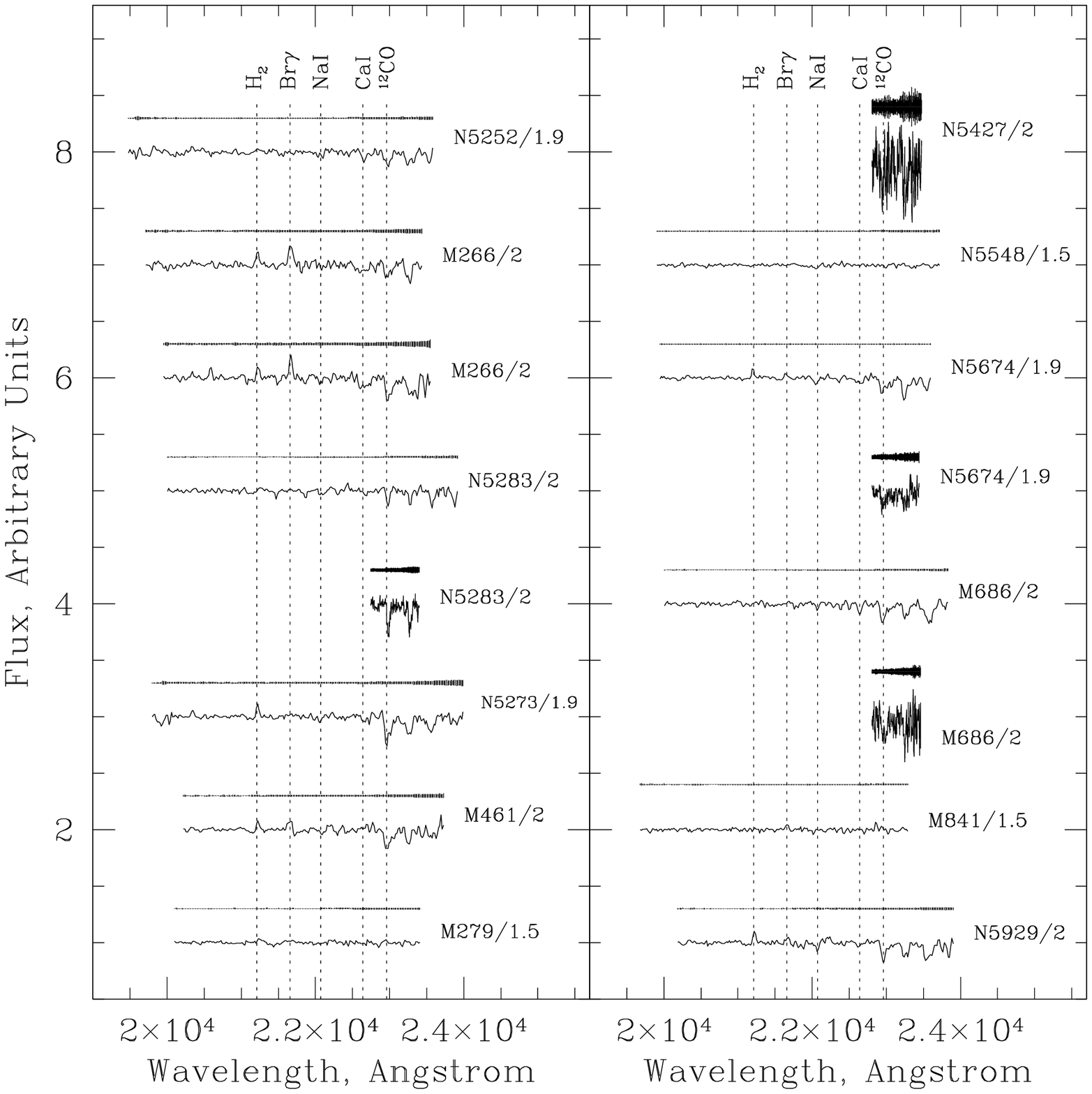}
\clearpage
\plotone{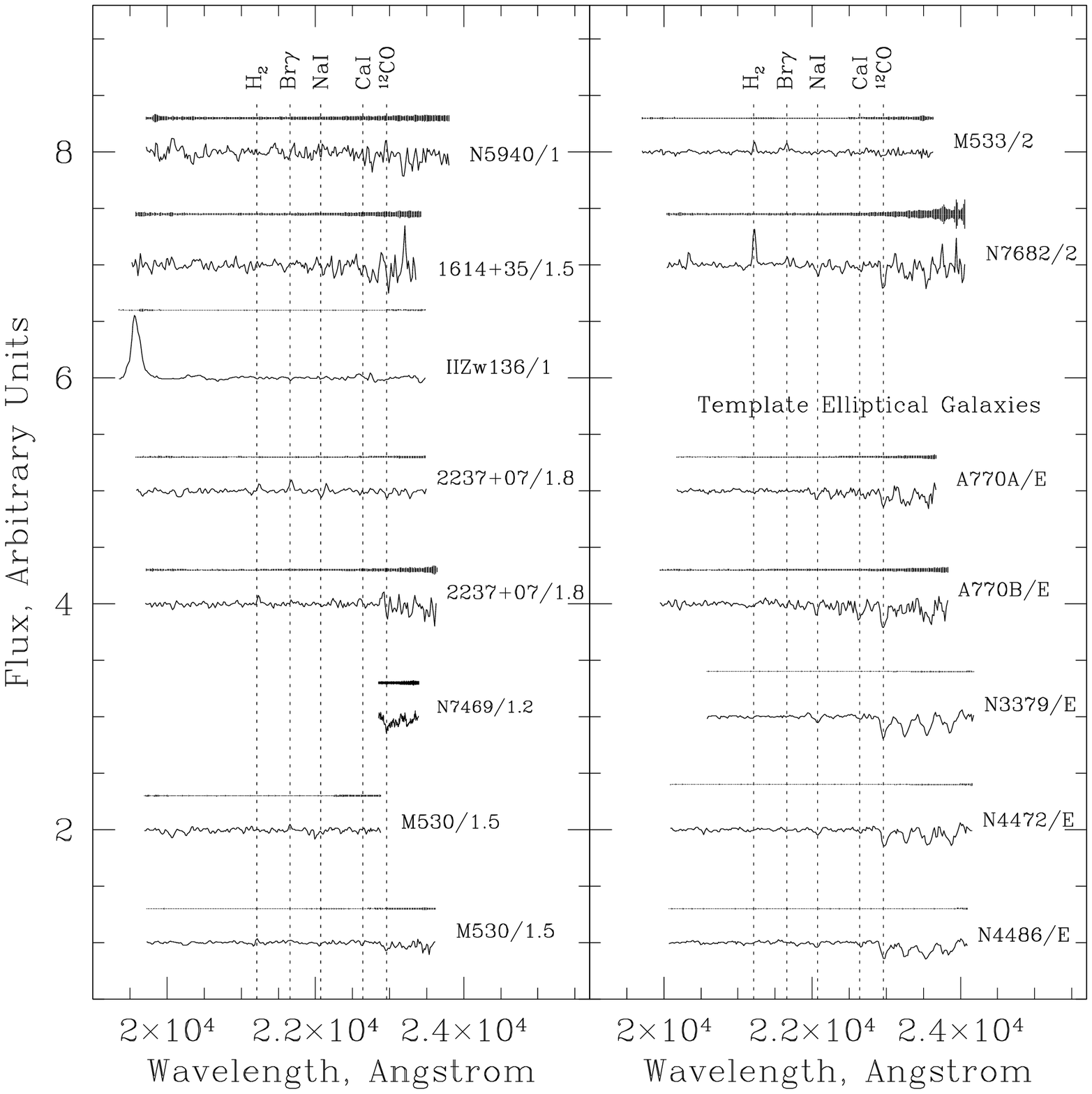}
\clearpage
\plotone{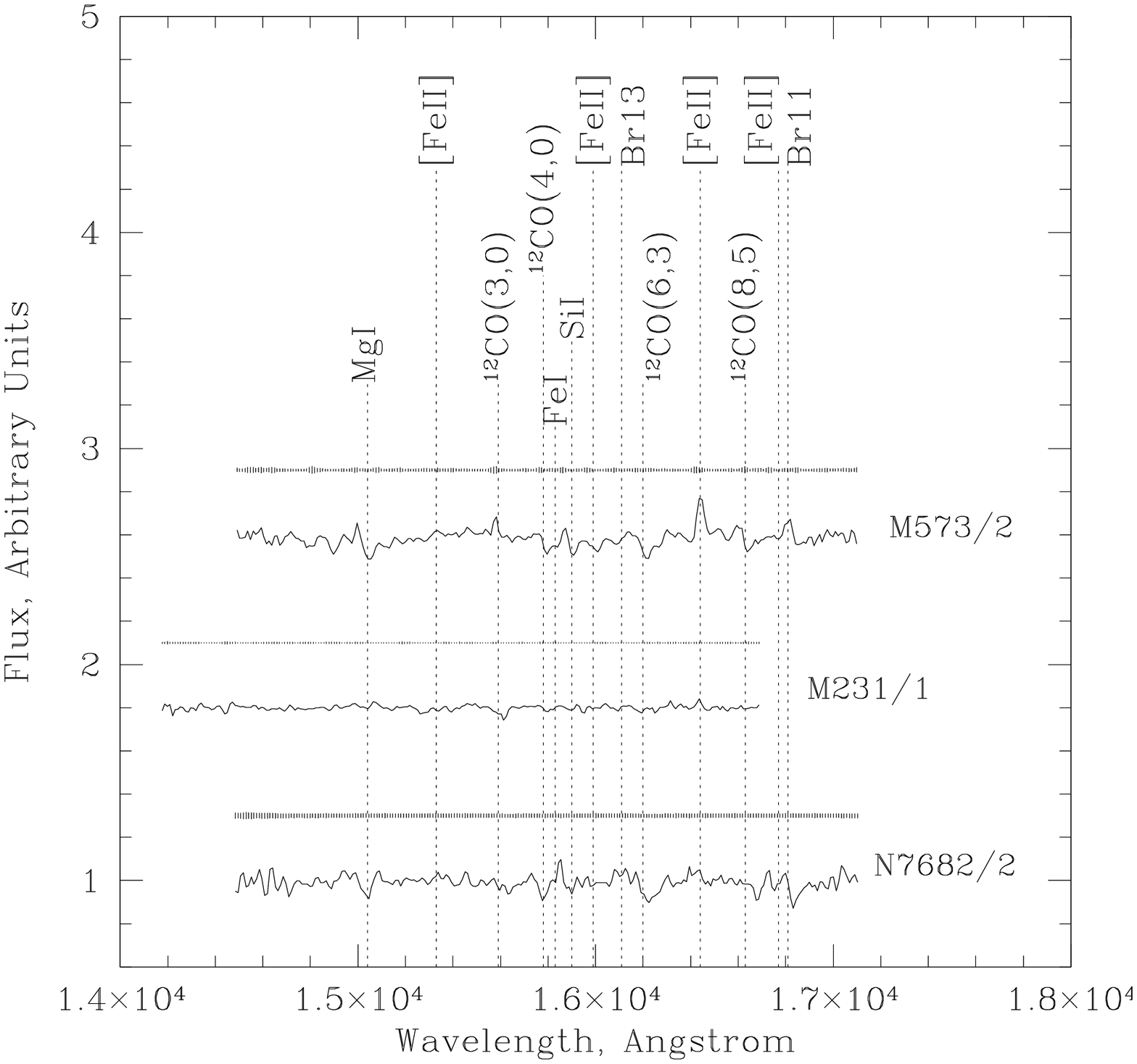}
\clearpage
\plotone{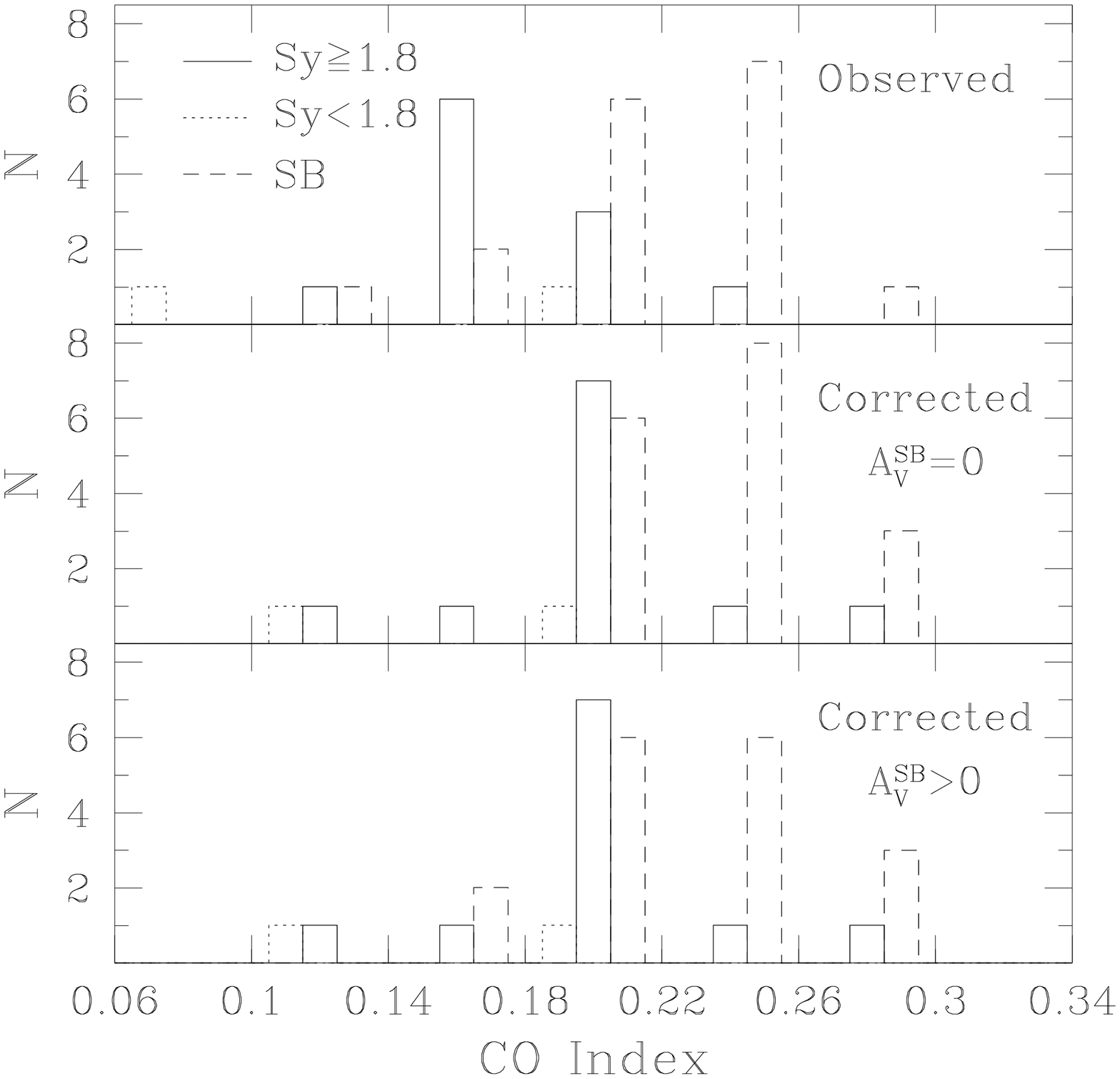}
\clearpage
\plotone{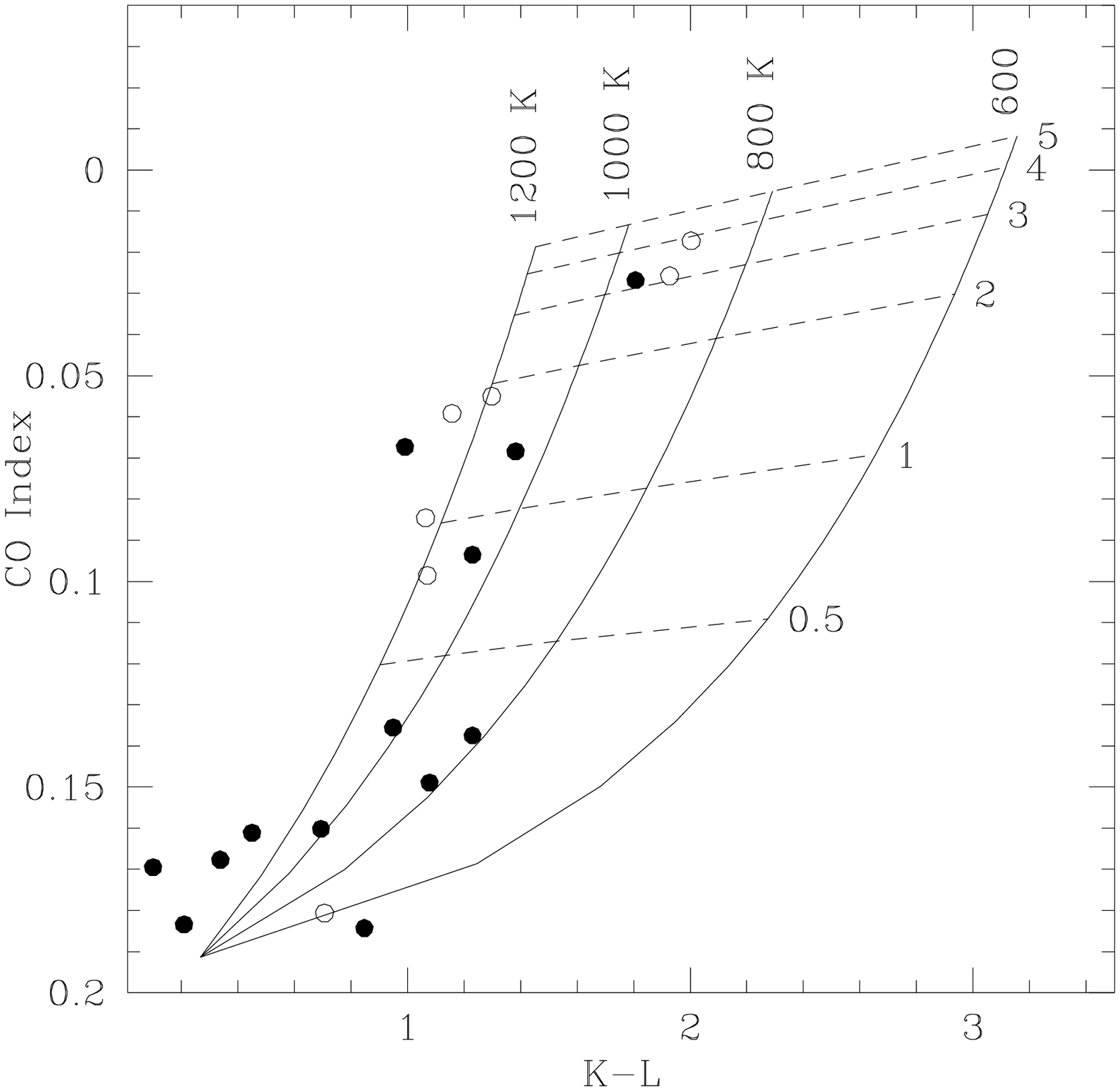}
\clearpage
\plotone{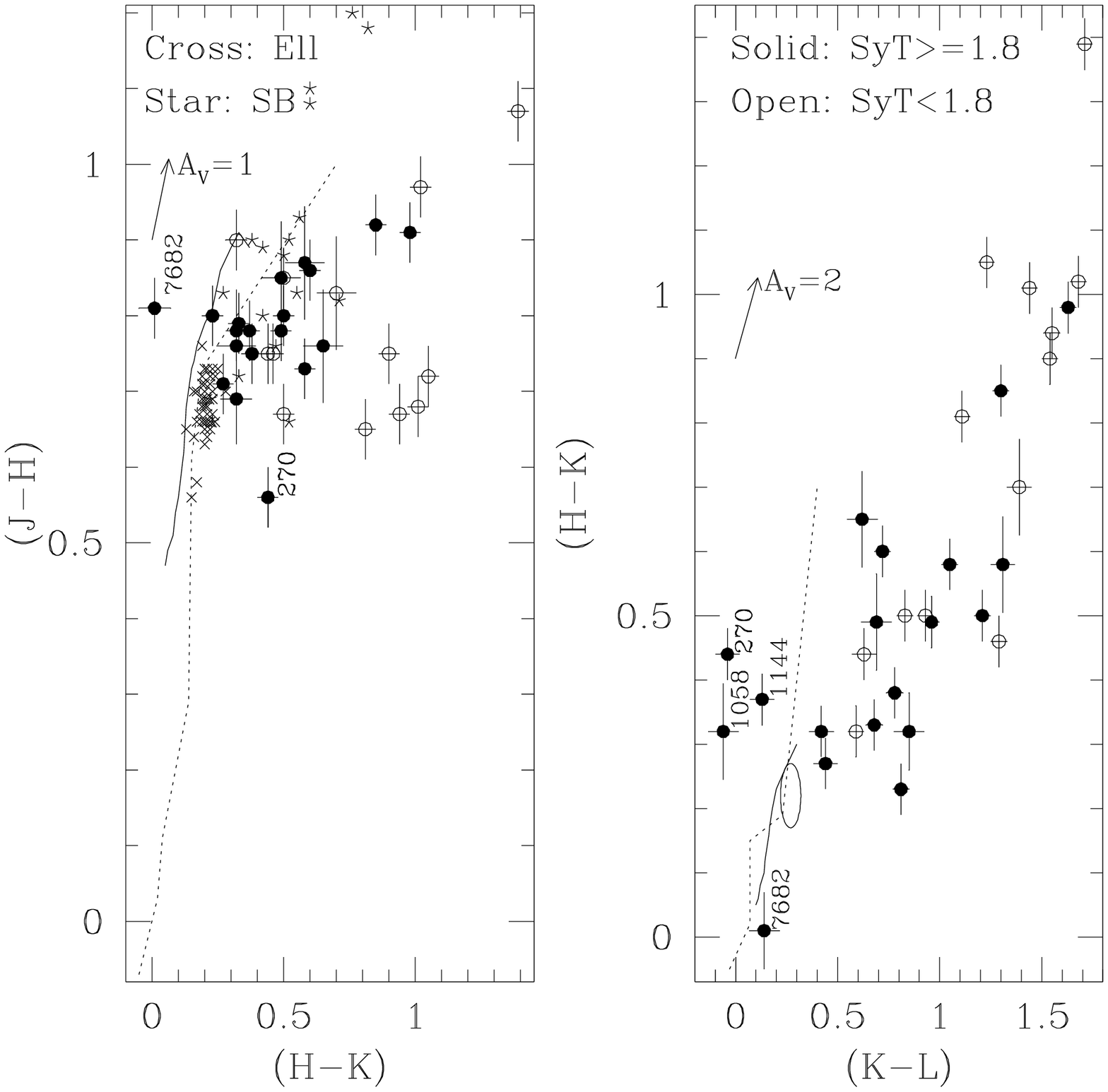}

\end{document}